\documentclass[namedreferences]{SolarPhysics}
\usepackage[hyperref,optionalrh,solaromanenum]{spr-sola-addons} 
\usepackage{graphicx}                    
\usepackage{amssymb}                    
\usepackage{color}                       
\usepackage{breakurl}                         

\newcommand{\RA}[1]{#1} 

\begin{document}

\begin{article}

\begin{opening}


\title{The Effect of ``Rogue'' Active Regions on the Solar
Cycle}

%
 \author[addressref={ELTE}
    ]{\inits{M.}\fnm{Melinda}~\lnm{Nagy}}
 \author[addressref={UdeM,college}
    ]{\inits{A.}\fnm{Alexandre}~\lnm{Lemerle}}
 \author[addressref={UdeM}
    ]{\inits{F.}\fnm{Fran\c{c}ois}~\lnm{Labonville}}
 \author[addressref={ELTE}
    ]{\inits{K.}\fnm{Krist\'of}~\lnm{Petrovay}}
 \author[addressref={UdeM}
    ]{\inits{P.}\fnm{Paul}~\lnm{Charbonneau}}

%
\runningauthor{M. Nagy \emph{et al.}}
\runningtitle{The Effect of ``Rogue'' Active Regions on the Solar Cycle}

\address[id=ELTE]{Department of Astronomy, Institute for Geography and Earth Sciences, E\"{o}tv\"{o}s University, Budapest, Hungary}
\address[id=UdeM]{D\'epartement de Physique, Universit\'e de Montr\'eal, Montr\'eal, QC, Canada}
\address[id=college]{Coll\`ege de Bois-de-Boulogne, Montr\'eal, QC, Canada.}

\begin{abstract}

\RA{
The origin of cycle-to-cycle variations in solar activity is currently
the focus of much interest. It has recently been pointed out that large
individual active regions with atypical properties can have a significant impact on the long term behaviour of solar activity. We investigate this possibility in more detail
using a recently developed 2$\times$2D dynamo model of the solar magnetic
cycle. We find that even a single ``rogue'' bipolar magnetic region (BMR)
in the simulations can have a major effect on the further development
of solar activity cycles, boosting or suppressing the amplitude of
subsequent cycles. In extreme cases an individual BMR can completely
halt the dynamo, triggering a grand minimum. Rogue BMRs also have the potential to
induce significant hemispheric asymmetries in the solar cycle.  To
study the effect of rogue BMRs in a more systematic manner, a series of
dynamo simulations were conducted, in which a large test BMR was manually
introduced in the model at various phases of cycles of different
amplitudes. BMRs emerging in the rising phase of a cycle can modify the
amplitude of the ongoing cycle while BMRs emerging in later phases will
only impact subsequent cycles. \RA{In this model,} the strongest impact on the
subsequent cycle occurs when the rogue BMR emerges around cycle
maximum at low latitudes but the BMR does not need to be strictly
cross-equatorial.  Active regions emerging as far as $20^{\circ}$
from the equator can still have a significant impact. We demonstrate that the combined effect of the magnetic flux, tilt angle and polarity separation of the BMR on the dynamo is \emph{via} their
contribution to the dipole moment, $\delta D_{\mathrm{BMR}}$. Our results indicate
that prediction of the amplitude, starting epoch and duration of a cycle requires an accurate
accounting of a broad range of active regions emerging in the previous cycle.

}
\end{abstract}

%

\end{opening}

\section{Introduction}
The significantly lower amplitude of the ongoing Solar Cycle 24 in
comparison to previous cycles has prompted increased interest in the
origin of cycle-to-cycle variations in solar activity. Experience has
shown that the best candidate for a physical precursor of the
amplitude of an upcoming cycle is the peak strength of the solar polar
magnetic fields (or alternatively, the solar dipole moment), reached
typically around the time of solar minimum (\citealt{Svalgaard2005};
\citealt{Petrovay2010}; \citealt{Petrie2014}).

The critical issue still open is how, in turn, the amplitude of the
dipole field is determined by solar activity in the previous cycle. In
the currently most widely discussed flux transport dynamo scenario,
the polar fields are built up by the poleward transport of
$f$-polarity magnetic flux from active regions. This transport is
mainly due to meridional circulation, so variations in meridional
circulation have been invoked as a key factor in inter-cycle activity
variations (\citealt{Dikpati2010}, \citealt{Jiang2010}, \citealt{Hathaway2014},
\citealt{Upton2014}).

An alternative possibility has been highlighted by
\citet{Cameron2010}  who stressed the importance of the tilt angle
$\alpha$ of bipolar active regions relative to the east-west
direction.\footnote{Throughout this paper, $\alpha$ is taken to
increase in the clockwise direction. For normally oriented active
regions obeying Joy's law, $\alpha$ is then positive on the N
hemisphere and negative on the S hemisphere.} Clearly, for a
$\alpha=0$, $f$- and $p$-polarity flux would be transported towards
the poles in equal rates, resulting in no net change in the polar
flux. An increasing tilt angle will then lead to an increasing polar
field strength. This opens two intriguing possibilities. On the one
hand, tilt angles
\RA{are known to be anticorrelated to cycle
amplitude (\citealt{Dasi2010}). The origin of this anticorrelation may
be related to the dynamics of the emerging flux loop or to the
meridional inflows towards the active latitude zone associated with
the torsional oscillation pattern, the amplitude of which is
determined by the level of solar activity. This ``tilt quenching'' is}
an important nonlinear feedback effect of solar
activity level on the tilt angles, and thereby on the buildup of polar
fields that will serve as seed fields for the next cycle.

On the other hand, a random scatter of tilt angles around the mean
value determined by the above process will introduce some degree of
stochasticity in the process. In particular, a few ``rogue'' active
regions disobeying Joy's law or Hale's polarity rule
(\citealt{McClintock2014}, \citealt{Jiang2014}) can potentially play
havoc with the buildup of polar fields, especially if they emerge near
the equator. The theoretical background of this, as outlined by
\citet{Cameron2015}, is that the net solar dipole moment can only be
affected by active regions emerging fully in one hemisphere if one of
their polarities is preferentially cancelled between the two
hemispheres by diffusion across the equator. This effect is stronger
for ARs emerging near the equator. Furthermore, the odd AR emerging
{\it across} the equator contribute directly to the solar dipole
moment without invoking the necessity of cross-equatorial magnetic
diffusion. In such cases, separation of the two polarities is aided by
the
poleward surface meridional flow.
As the total magnetic flux in the polar cap is comparable to
the magnetic flux in a single large active region
(\citealt{WangSheeley1991}), even one AR emerging near (or, indeed,
across) the equator can have a major distorting effect of polar flux
buildup. The effect of rogue ARs at higher latitudes is, however, much
less dramatic (\citealt{Yeates2015}).

All this suggests a scenario where intercycle variations are normally
governed by the nonlinear feedback effect on the tilt angles, offering
a chance to forecast the amplitude of upcoming cycles in many cases,
while once in a while rogue AR introduce large statistical
fluctuations in the process, leading to unpredictable variations in
the level of solar activity. One may speculate that in extreme cases
such freak events may even trigger longer episodes of unusually low or
unusually high activity, \emph{i.e.} grand minima or grand maxima.

Pure mean-field dynamo models are inadequate for the study of such
effects. Instead, individual active regions, manifest in the models as
``bipolar magnetic regions'' (BMRs) must be incorporated in the models
in a way that correctly reproduces their statistical mean
characteristics while allowing for stochastic variations.  One such
dynamo model, the ``2$\times$2D'' dynamo model has recently been developed by
\citet[hereafter L2015 and L2017, respectively]{Lemerle2015,Lemerle2017}.
The purpose of the present paper is to study
the effect of individual BMR, with special regard to rogue BMR on
intercycle variations in the 2$\times$2D dynamo model. An added advantage of
this study is that it allows us to consider extremely long time
series, covering hundreds of activity cycles. In contrast, the
observational record of solar activity at our disposal with sufficient
details regarding the sizes, positions and magnetic polarities of the
spots only covers about half a dozen solar cycles, with less
detailed data extending back to about 20 solar cycles.

The structure of the paper is as follows. Section 2 briefly recalls
the main features of the 2$\times$2D dynamo model used. In Section 3
we discuss salient features of the solutions with respect to
stochasticity and cycle-to-cycle variations, highlighting the role of
rogue BMRs. Section 4 discusses the most extreme effects of rogue BMR
citing a few cases where such BMR naturally arise within the
2$\times$2D dynamo model. In Section 5 a systematic series of
numerical experiments are performed wherein a large ``test'' BMR
departing strongly from Joy's law
is manually introduced in the simulation at various times, locations and with various other parameters,
and the effects on the magnetic cycle are analyzed.
Section 6 concludes the paper.

\section{The Model Used}

The \citet{Lemerle2017} solar cycle model couples a
conventional surface flux transport (SFT) 2D simulation defined over a
spherical surface (see L2015) to an equally conventional 2D
axisymmetric Flux Transport Dynamo (FTD) simulation defined in a
meridional plane (\citealt{Charbonneau2005}). In the resulting hybrid
kinematic 2$\times$2D Babcock-Leighton dynamo model, the SFT
component provides the surface poloidal source term for the FTD
simulation through the latter's surface boundary condition, while the
FTD component provides the magnetic emergence events required as input to
the SFT simulation.

It should be noted that other attempts at extending the mean field
dynamo models to 3D with the inclusion of individual active regions do
exist (\citealt{Yeates2013}, \citealt{Miesch2014}). The main advantage
of the 2$\times$2D model lies in its numerical efficiency and in the
fact that it has been carefully calibrated to resemble the actual Sun.
As the model includes a complete latitude-longitude representation of
the simulated solar surface, the impact of varying characteristics of
emerging BMRs (flux, pole separation, tilt angle, \emph{etc.}) can be
accounted for; as it does not solve the induction equation in three
spatial dimensions, relatively high spatial resolution can be achieved
within each model component, while allowing simulations extending over
many thousands of simulated cycles. Detailed description of the model
components can be found in the afore-cited papers (and especially
L2015). In this section we merely provide an overview of the features
that are most relevant to the investigations described futher below.

The physical conditions driving the BMR emergences from the convective
zone and their further temporal evolution in the solar photosphere are
well described by the MHD induction equation:

\begin{equation}\label{eq:MHD}
    \frac{\partial\mathbfit{B}}{\partial t} = \nabla\times
    \left( \mathbfit{u}\times\mathbfit{B} - \eta\nabla\times\mathbfit{B} \right).
\end{equation}

\noindent The large-scale flow $\mathbfit{u}$
includes contributions from meridional circulation
and differential rotation. The former is described by a modified form of the
flow profile of \citet{vanBallegooijen1988}, while for the latter we
adopt the helioseismically-calibrated parametric form of
\citet{Charbonneau1999}. Both of these are considered axisymmetric
and steady, as per the kinematic approximation.
The total magnetic diffusivity $\eta$ in the internal dynamo simulation
follows the parametric profile $\eta(r)$ of \citet{Dikpati1999}. In the SFT
component we set $\eta_R \approx 10^{12}-10^{13}\mathrm{cm}^2\mathrm{s}^{-1}$,
reflecting the strong effective diffusive transport provided
by the supergranular flow.

The axisymmetric magnetic field simulated in the FTD component of the model
is expressed as:
\begin{equation}\label{eq:B_FTS}
    \mathbfit{B}(r,\theta,t) =
    \nabla\times[A_{\phi}(r,\theta,t)\hat{\mathbf{e}}_{\phi}]
    + B_{\phi}(r,\theta,t)\hat{\mathbf{e}}_{\phi},
\end{equation}
\noindent where $\mathbfit{B}_{\mathrm{P}}= B_r\hat{\mathbf{e}}_r +
B_{\theta}\hat{\mathbf{e}}_{\theta}$ and
$B_{\phi}\hat{\mathbf{e}}_{\phi}$  are the poloidal and toroidal
vector components of the field, respectively.  Substitution into
Equation \ref{eq:MHD} leads to two evolution equations for the scalar
components $A_{\phi}(r,\theta,t)$ and $B_{\phi}(r,\theta,t)$:
\begin{eqnarray}\label{eq:timeFTS}
        \frac{\partial A_{\phi}}{\partial t} &=& -\frac{1}{\varpi}(\mathbfit{u}_{\mathrm{P}}\cdot\nabla)(\varpi A_{\phi}) + \eta\left(\nabla^2 - \frac{1}{\varpi^2}  \right)A_{\phi}, \\
  \frac{\partial B_{\phi}}{\partial t} &=& -\varpi(\mathbfit{u}_{\mathrm{P}}\cdot\nabla) \left(\frac{B_{\phi}}{\varpi}\right) + \eta\left(\nabla^2 - \frac{1}{\varpi^2}  \right)B_{\phi} \nonumber \\
        &-& (\nabla\cdot\mathbfit{u}_{\mathrm{P}})B_{\phi} + \frac{1}{\varpi} \frac{\partial \eta}{\partial r} \frac{\partial (\varpi B_{\phi})}{\partial r} + \varpi\mathbfit{B}_{\mathrm{P}}\cdot\nabla\Omega,
\end{eqnarray}
\noindent where $\varpi = r\sin\theta$,
$\mathbfit{u}_{\mathrm{P}}$ is the meridional flow and
$\Omega(r,\theta)$ is the angular velocity.\\
The magnetic field in the SFT code is considered radial, hence only
the $r-$component of Equation \ref{eq:MHD} is solved at $r = R$:
\begin{eqnarray}\label{eq:timeSFT}
  \frac{\partial B_R}{\partial t} &=& - \frac{1}{R\sin\theta}\frac{\partial}{\partial\theta} \left[ u_{\theta}(R,\theta)B_R\sin\theta \right] - \Omega(R,\theta)\frac{\partial B_R}{\partial \phi} \nonumber \\
  &+& \frac{\eta_R}{R^2} \left[ \frac{1}{\sin\theta}\frac{\partial}{\partial\theta} \left( \sin\theta \frac{\partial B_R}{\partial \theta} \right) + \frac{1}{\sin^2\theta}\frac{\partial^2 B_R}{\partial \phi^2} \right] \nonumber \\
  &-&\frac{B_R}{\tau_R} + S_{\mathrm{BMR}}(\theta,\phi,t).
\end{eqnarray}
The linear sink term $-B_R/\tau_R$ allows for exponential decay of the
surface field (\citealt{Baumann2004}, \citealt{Baumann2006}),
mimicking subduction of the polar cap magnetic field by
the meridional flow.
The other additional term $S_{\mathrm{BMR}}(\theta,\phi,t) =
\sum_iB_i(\theta,\phi)\delta(t-t_i)$ is the source of new BMR
emergence events with $\delta$ the Dirac delta. The internal module receives
the longitudinally averaged SFT solution $\langle B_R
\rangle(\theta,t)$ at every FTD time step. This step provides the
upper boundary condition. The coupling from the FTD towards the SFT
simulation is the emergence of new BMRs. This is based on a
semi-empirical emergence function $F_B(\theta,t)$ that gives the
probability that the emergence of an active region will occur, given
the internal distribution of magnetic fields within the FTD component
of the model. Whenever one such emergence takes place,the
characteristics of the emerging bipolar magnetic region ---flux, pole
separation, tilt angle--- are randomly drawn from distribution
functions for these quantities built from observed statistics of solar
active regions (see Appendix A in L2015).

The emergence function
incorporates a field strength threshold, below which the emergence
probability falls rapidly to zero.
This implies that the dynamo is
not self-excited, in that it cannot amplify a seed magnetic field
if the strength of the latter lies below this threshold.
Along with this
lower operating
threshold, the only
other nonlinearity introduced in the model is a reduction of the
average active region tilt angle ($\alpha$) with internal magnetic
field strength:

\begin{equation}\label{eq:tiltquenching}
    \alpha_q = \frac{\alpha}{1 + (B_{\phi}/B_q)^2}
\end{equation}
where $B_q$ is the quenching field amplitude. Although this specific
mathematical form is largely \emph{ad hoc}, the idea of tilt quenching is
supported by simulations of the rise and emergence of thin flux tubes.
Physically, it reflects the fact that more strongly magnetized flux
tubes rise more rapidly through the convection zone, and suffer less
distorsion by the Coriolis force during their rise (see
\citealt{Fan2009}, and references therein).

\begin{figure}
  \centering
  \includegraphics[width=\textwidth]{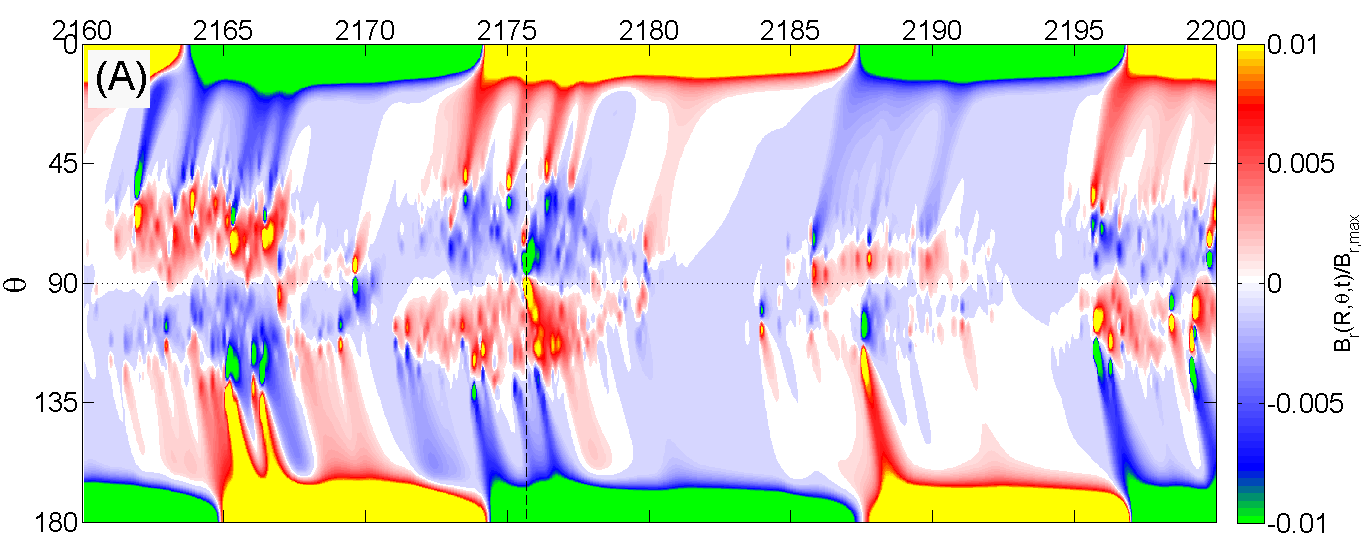}
  \includegraphics[width=\textwidth]{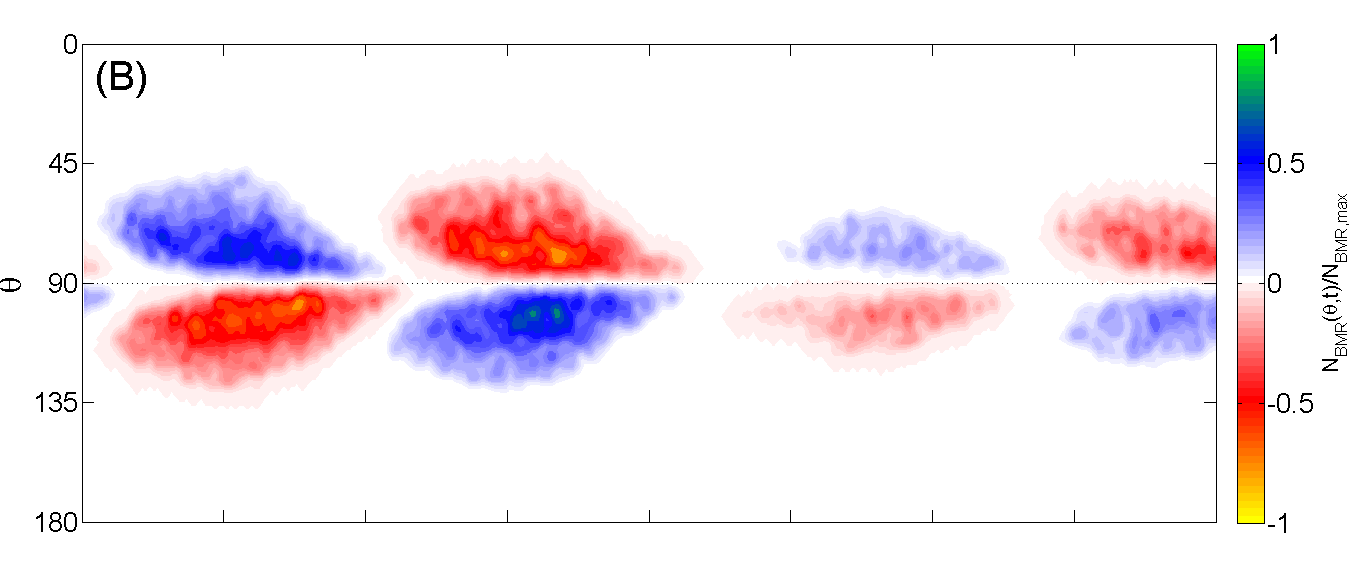}
  \includegraphics[width=\textwidth]{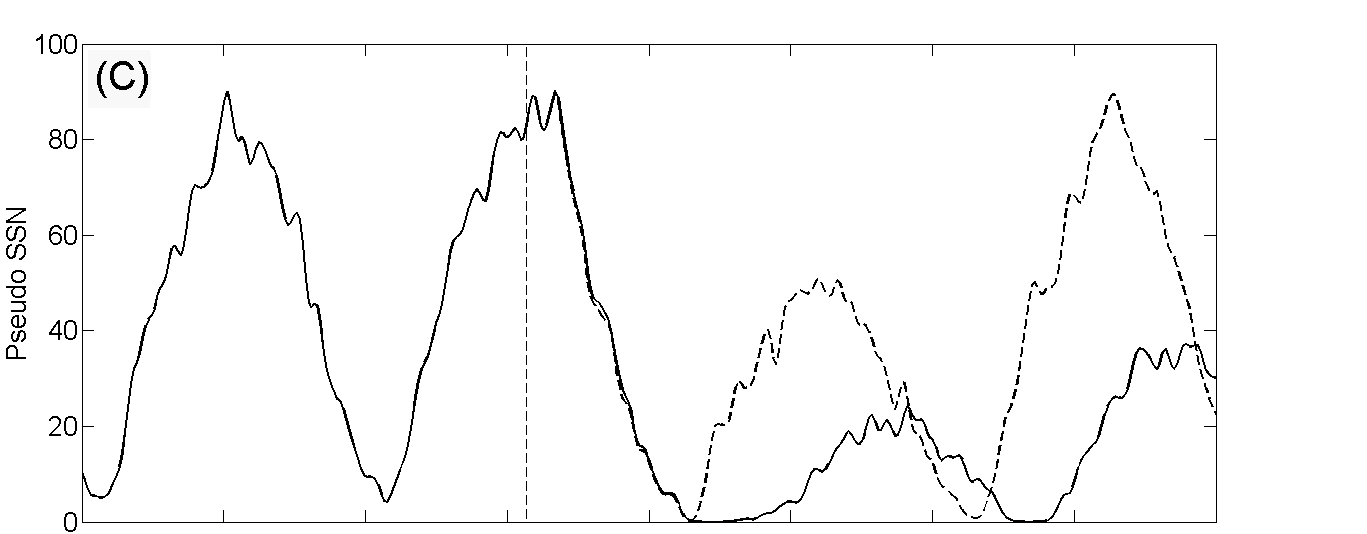}
  \includegraphics[width=\textwidth]{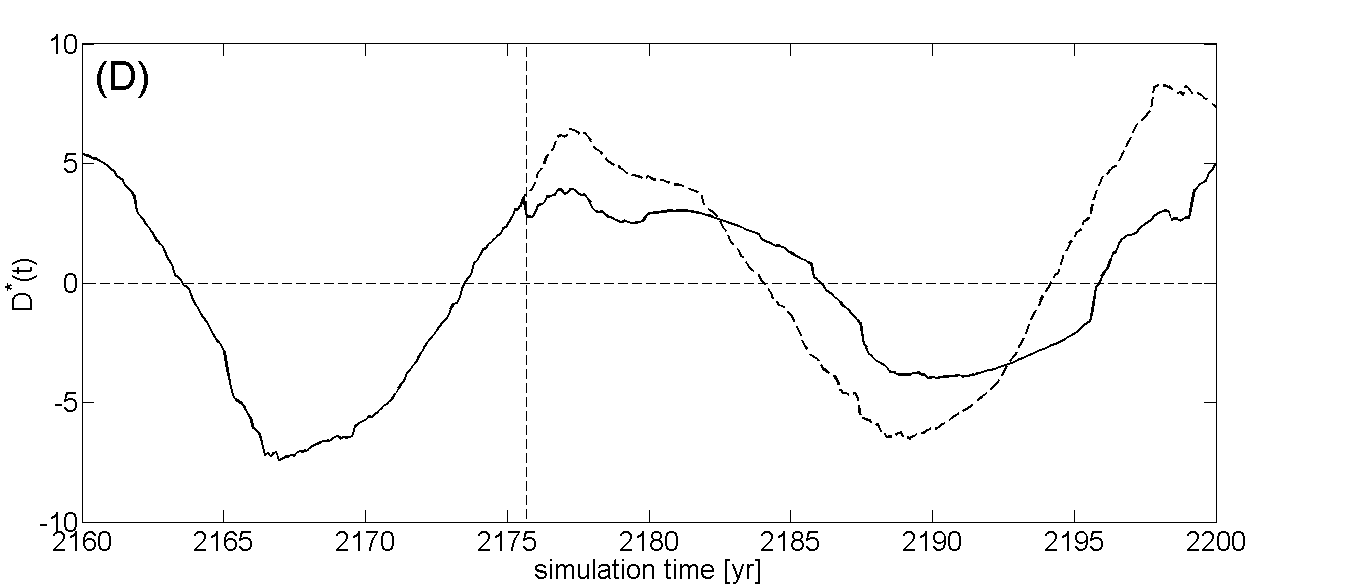}\\
  \caption{In panel (A) the surface time-latitude magnetogram is plotted for a representative segment of a reference simulation run. Panel (B) shows the butterfly diagram of BMR emergences, colored according to the trailing polarity,
for the same period of time. The pseudo-SSN (C) and dipole moment (D) plotted with solid lines correspond to the data shown in panels (A) and (B). The dashed lines on panels (C) and (D) result from artificially removing a single large BMR emergence from the simulation (see Section \ref{sect:stochast}). }\label{fig:thenumber}
\end{figure}

The reference solar cycle solution presented in L2017, which is
adopted in the numerical experiments carried out in the present work,
is defined by 11 adjustable parameters, which were optimized using a
genetic algorithm designed to minimize the differences between the
spatiotemporal distribution of emergences produced by the model, and
the observed sunspot butterfly diagram. Figure \ref{fig:thenumber}
presents a 40-yr long representative segment of this reference
simulation. Panels (A) and (B) show time-latitude plots (butterfly
diagrams) for the zonally-averaged surface magnetic field and for BMR
emergence locations, respectively.

We note that emergence at low latitudes only results from the
interplay between the internal magnetic field distribution within the
FTD component and the emergence function. We do not impose a mask to
restrict BMR emergences to low heliographic latitude, although our
emergence function does include a high latitude cutoff at
$\pm70^{\circ}$ latitude, as suggested by stability analyses of thin flux
tubes below the solar convection zone (see L2015 for further
discussion).

It should also be noted that due care is warranted to ensure magnetic flux
conservation within the SFT component of the model to a high level
of numerial accuracy. This is because the polar cap flux, feeding
back from the STF into the FTD model component, accounts for less
than one percent of the total emerging (unsigned) BMR flux in the course
of a typical cycle. Numerically accurate reproduction of the dipole
moment to a relative accuracy of $10^{-3}$ (say) then requires
numerical flux conservation at better than the $10^{-5}$ level
(see L2017).

A pseudo-sunspot number time series can be constructed by simply
summing the number of emergences taking place over subsequent one-day intervals
in the simulation. Because we do not distinguish between individual spots
and groups in the simulation, we simply rescale the amplitude of this
time series to yield values commensurate with the international sunspot number.
The resulting time series, smoothed with a 13-month running boxcar filter,
is plotted as the black solid line
in panel (C) of Figure~\ref{fig:thenumber} for the same simulation segment
as on the upper two panels.

The axial dipole characterizing the SFT component is readily
computed as:
\begin{equation}\label{eq:dipolemoment}
    D^*(t) = \frac{D(t)}{R^2} = \frac32 \int_0^{\pi}\langle B_R \rangle^{\phi}(\theta,t)\cos\theta\sin\theta \mathrm{d}\theta,
\end{equation}
that is also plotted in panel (D) of Figure~\ref{fig:thenumber}. As
shown in L2017 Section 4.2 and in agreement with the Sun, the magnitude
of this dipole correlates very well (linear correlation coefficient
$r\approx 0.9$) with the amplitude of the subsequent (pseudo-) SSN
cycle amplitude, but not with the amplitude of the cycle during which
the dipole is building up ($r\approx 0.25$). This indicates that the
primary stochasticity driving cycle amplitude fluctuations operates in
the course of the dipole buildup.
This is readily seen on the bottom panel of Figure~\ref{fig:thenumber}:
the first two pseudo-SSN cycles have about the
same amplitude of pseudo-sunspot numbers, but the dipole moments they
generate in their descending phase differ by a factor of two. In
contrast, the second and third pseudo-SSN cycles \RA{(solid black curve)} differ by a factor of four in
amplitude, yet generate dipoles peaking at almost exactly the same
strength. This indicates that, in this simulation model (and most
likely in the Sun as well as), it is not only the number of BMRs that
sets the strength of the dipole moment, but rather the combination of
physical parameters (polarity separation, tilt, \emph{etc}.) of the emerging
active regions. We now turn to a more detailed analysis of this
behavior and its consequences for cycle predictability.

\section{Stochasticity and Predictability}\label{sect:stochast}

The L2017 solar cycle model is a kinematic dynamo, in the
sense that the large-scale flows introduced therein ---differential
rotation and meridional circulation--- are assumed steady. This
implies that the transport of the poloidal component from the surface to
the deep interior as well as its shearing by internal differential rotation,
and therefore the poloidal-to-toroidal part of the dynamo loop, are
fully deterministic processes; any variation in the internal toroidal
field so generated can only arise from a corresponding variation of
the surface dipole. The production of the latter, on the other hand,
is strongly influenced by the specificities of active region emergences:
their magnetic flux, pole separation, tilt angle with respect to the E-W
direction, \emph{etc}. This represents a source of stochastic fluctuations,
which in fact dominates the cycle-to-cycle variations produced by the
model.

That this should be the case is not {\it a priori} obvious, and reflects
a specificity of the global surface magnetic flux
budget of the Sun. As pointed out by \citet{WangSheeley1989}, for a typical solar activity cycle
the observed polar cap flux at the time of peak dipole strength
is a few $\approx 10^{22}\,$Mx; this
is similar to the unsigned flux in one pole of a single
large active region, and roughly one percent of the total unsigned
flux emerging in active regions in the course of a typical solar activity
cycle. Accurate modelling of the surface magnetic flux evolution
thus requires that all properties of emerging active regions
be known accurately if the buildup of the dipole in the descending
phase is to be modeled (or predicted) with good accuracy.

\begin{figure}
\centering
\includegraphics[width=0.8\textwidth]{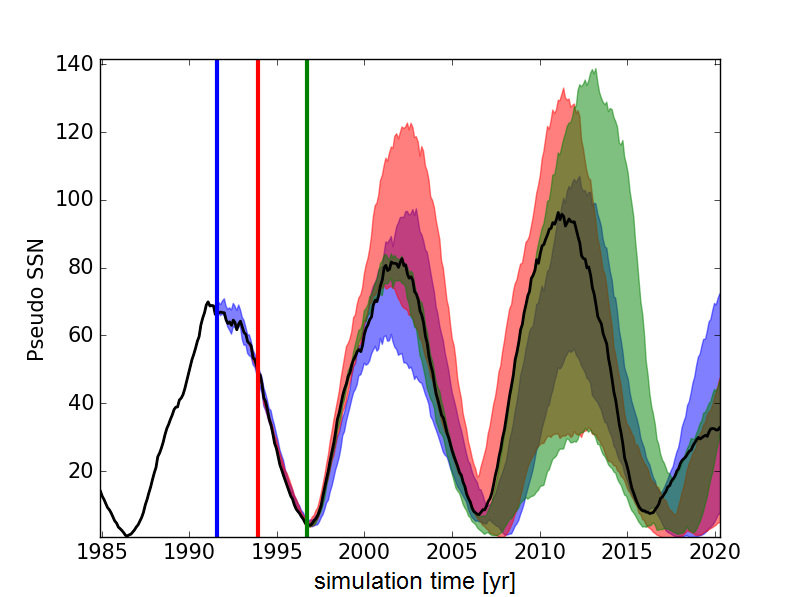}\\
\caption{Divergence of the pseudo-SSN time series in the
L2017 solar cycle model. The black curve
is a reference time segment for a single run, and the colored
bands give the min/max range for three ensembles of five solutions
in which the random number generator controlling the statistics
is reset at the time indicated by the correspondingly colored
vertical line segments. In all cases divergence typically sets in
approximately one pseudo-SSN cycle period after this reset.
}
\label{fig:experiment0}
\end{figure}

This sensitivity can be readily illustrated by the following
numerical experiment, the results of which are displayed on
Figure \ref{fig:experiment0}. The black solid line is a time series segment
of the smoothed monthly pseudo-sunspot number produced
by a representative simulation run of the L2017 model.
Now suppose that this reference run is the ``real'' sunspot
number time series, and one is attempting to use the model
to forecast future cycles
by running ensembles of Monte Carlo simulations
in which the characteristics of active regions are drawn
randomly from empirically constructed distributions for the
relevant quantities.
The three semi-transparent colored bands are the min/max ranges
in three such ensembles of five simulations in which, at a time
indicated by the correspondingly colored vertical line segments,
the random number generator controlling the draw of numerical
values for magnetic fluxes,
tilt angle, \emph{etc}., for the emerging active regions, is reset to
another seed value. All model parameters remain the same, so that
the diverging subsequent evolutions reflect only the different
statistical realizations of the active region emergence process.

In all cases, a significant time delay between the reset and the
divergence of the associated pseudo-SSN time series is observed.
The extent of this time delay reaches a full cycle period when
the reset takes place at minimum activity (green band), with the
reset solutions tracking the original very well all the way
to the subsequent minimum.
If on the
other hand the reset takes place at solar maximum (blue) or during the
first half of the descending phase (red), divergence
sets in already during the rising phase of the subsequent cycle,
and the resulting range of cycle peak amplitudes ends up being a
large fraction of the peak amplitude for the unperturbed
reference solution (black solid line). The largest prediction window
is thus possible working off cycle minimum, in which case the
amplitude and duration of the subsequent cycle are well and
reliably reproduced.
Beyond the cycle following the reset of the emergence statistics
(here from $t=2006$ onward),
all solutions diverge strongly, with
even the timing of pseudo-SSN minima and maxima being affected even
though the meridional flow speed remains the same in all cases.

Evidently,
beyond about one cycle period the solutions diverge to the point
of making any long-term prediction a futile undertaking, at least
in the Monte Carlo form implemented here in which active region
emergent properties are drawn randomly from empirically constructed
statistical distributions.
Nonetheless, in all these solutions,
once the surface dipole is built at the end of
a cycle's descending
phase, the evolution of the subsequent cycle
---including peak amplitude, timing of maximum, duration, \emph{etc}.---
is set.
This is the case in the Lemerle \emph{et al.} model, unavoidably so in view
of its aforementioned kinematic
nature. It also seems to be the case in the Sun, as
indicated by the fact that the cycle prediction methods based on measures
of the solar surface dipole at times of activity minimum
(\emph{e.g.} \citealt{Schatten1978}; \citealt{Svalgaard2005}; \citealt{Choudhuri2007}), whatever their specific implementation may be,
tend to fare better than most, and have done so
again with the unusually low sunspot Cycle 24 (for a review see \citealt{Petrovay2010}).

In this context, any hope of extending the predictability window
beyond one cycle period clearly lies with an accurate prediction of
the dipole moment building up in course of a cycle, and understanding
which characteristics of emerging active regions have the most
potential impact on this process. Put differently, we need to
understand which active regions have the potential of completely
derailing a prediction scheme based on assimilation of emerging active
region data in the course of a cycle \citep{Upton2014}. The simulation segment plotted
on Figure \ref{fig:thenumber} offers an extreme ---through not
unreasonably so--- specific example. Consider again the pseudo-SSN
time series (black solid line in panel (C)), the corresponding time
series for the surface dipole moment (panel (D)) and surface
magnetogram (panel (A)) in Figure \ref{fig:thenumber}. Here the cycle
peaking at year 2176 is followed by an extended minimum phase leading
to a subsequent cycle of much reduced amplitude, qualitatively
resembling the behavior observed with sunspot Cycles 23 and 24.  In
this specific simulation, at the time indicated by the vertical dashed
line segment, near cycle maximum a very large ($\approx 10^{23}\,$Mx, see
Table \ref{tab:BMRs}, 2nd column)  ``rogue'' active region emerged
very close to the equator, and with the trailing polarity closer to
the equator (anti-Joy tilt angle). This immediately leads to a
significant drop in the dipole moment, which ``stalls'' what was up to
then a typical buildup of the dipole.
Moreover, the low amplitude, Solar Cycle-24 like cycle, beginning at $t\approx 2182$ is
clearly also quite asymmetric as seen in the butterfly diagram, panel
(B) of Figure \ref{fig:thenumber}. In our example, the new active
regions appear first in the south, while in the north they appear a
few years later only. This suggests that the emergence of the rogue
BMR also affected the hemispheric asymmetry we observe during the next
cycle ---\emph{cf.} Section 4 below.

Consider now the artificial removal of this active region; this leads
to a subsequent evolution shown by the dashed lines, for the
pseudo-SSN, panel (C) and the dipole moment, panel (D). The large drop
in the dipole moment building up in the descending phase of the cycle
clearly has a strong impact on the subsequent cycle: upon removal, the
cycle begins its rising phase almost two years before it did in the
reference solution, and reaches a peak amplitude more than twice
higher and over two years sooner.

The $2.4\times 10^{23}\,$Mx flux of the rogue BMR appearing ---and
being removed--- in this simulation is indeed very high, but still not
unreasonably so in the solar context (\citealt{Tlatov2014};
\citealt{Munoz2015}; \citealt{Toriumi2017} ). During the descending
phase of the ongoing Cycle 24 a gigantic BMR emerged in October 2014.
This active region (AR2192) had a magnetic flux of about
$1.6\times10^{23}$Mx and it topped out at a size of 2750
micro-hemisphere. This was the biggest since AR6368 in November 1990
which had a size of 3080 micro-hemisphere, but even this was still
about half of the ``Great Sunspot of 1947'' which occupied about 6100
micro-hemisphere. Nonetheless, the cross-equatorial active regions
reported by \citet{Cameron2014} had fluxes of $2$--$3\times^{22}$ Mx
but still had huge effect on the dipole moment according to their
surface flux transport (SFT) simulations. The somewhat extreme example
of Figure \ref{fig:thenumber} is thus a perfect illustration of their
argument for potentially critical impact of emergences occurring close
to (or across) the solar equator, with the model providing the
additional benefit of self-consistently tracking the impact of such
emergences on the dynamo loop. Note also here how the $\approx 2\,$yr
cycle onset delay erased by the removal of this rogue active region
persists here through the last cycle plotted, even though this is a
kinematic model in which the average cycle period is set by the
meridional flow speed, which is held fixed throughout this whole
simulation.

From Figure \ref{fig:thenumber} it is clearly seen that after the
indicator line at $t \approx 2175.5$ the temporal evolution of the
dipole moments splits immediately. The separation happens due to the
removal of a peculiar or ``rogue'' active region identified in the
reference case. By ``rogue'' we refer to its extreme physical
properties such as high flux and angular separation. In addition, it
is also characterized by a high tilt angle in the opposite orientation to
what is expected from Joy's law. The properties of this active region
are listed in the second column of Table \ref{tab:BMRs}.

\begin{table}[h!]
  \begin{tabular}{r | l l l l l}
	$\theta_{\mathrm{lead}}$           & 84.8$^{\circ}$     & 82.5$^{\circ}$    & 95.6$^{\circ}$	& 112$^{\circ}$      & 89.5$^{\circ}$    \\
	$\theta_{\mathrm{trail}}$          & 89.5$^{\circ}$     & 86.3$^{\circ}$    & 104.1$^{\circ}$	& 118.6$^{\circ}$    & 82.1$^{\circ}$     \\
	$F $ [$10^{23}$ Mx]                & 2.43               & --7.04 	    & --3.58		& 4.39               & --1.39	         \\
	$\alpha$                           & $-8.65^{\circ}$    & $-10.92^{\circ}$  & $-15.53^{\circ}$  & --11.08$^{\circ}$   & 13.98$^{\circ}$    \\
	$d$                                & 31.64$^{\circ}$    & 20.16$^{\circ}$   & 32.11$^{\circ}$	& 34.80$^{\circ}$    & 30.97$^{\circ}$   \\
    $\delta D_{\mathrm{BMR}}$[$10^{23}$ Mx] & \RA{--0.2014} & \RA{0.4670  }     & \RA{0.5293 }      & \RA{--0.4633}      & \RA{--0.1810 }\\
    J/H                                & a-J/H              & a-J/H             & J/H               & J/H                & J/H, J/a-H\\
    \hline
	                          & Figure \ref{fig:thenumber} & Figure \ref{fig:shutdownDynamo}& Figure \ref{fig:restartDynamo} & Figure \ref{fig:asymmetry}
				  & Section \ref{sect:BMReffects}
  \end{tabular}
  \caption{Physical quantities of active regions discussed in the paper.
  Colatitudes $\theta_{\mathrm{lead}}$ and $\theta_{\mathrm{trail}}$ are the
  latitudinal positions of leading and trailing polarities;
  $F$ is the flux of the trailing polarity
  ($F_{\mathrm{trail}} = -F_{\mathrm{lead}}$); $\alpha$ is the tilt
angle and $d$ is the angular separation of leading and trailing
polarities. $\delta D_{\mathrm{BMR}}$, the contribution of the BMR to the global dipole moment, is defined according to Equation \ref{eq:thenumber}. J/H indicates whether the active region is (anti-)Joy/(anti-)Hale. In the case of the last column a J/H (J/a-H) test-BMR increases (decreases) the dipole moment during the experiments detailed in Section \ref{sect:BMReffects}. }\label{tab:BMRs}
\end{table}

\section{Extreme Effects of Peculiar Active Region Emergences}

In the case plotted in Figure~1 and discussed in the previous section,
the rogue BMR had a significant effect on the further development of
the activity cycles. This effect, however is still relatively moderate
compared to some truly extreme cases to be presented in what follows.

\subsection{Halt of Dynamo Action}\label{sect:dieDynamo}
    \begin{figure}[h!]
      \centering
      \includegraphics[width=\textwidth]{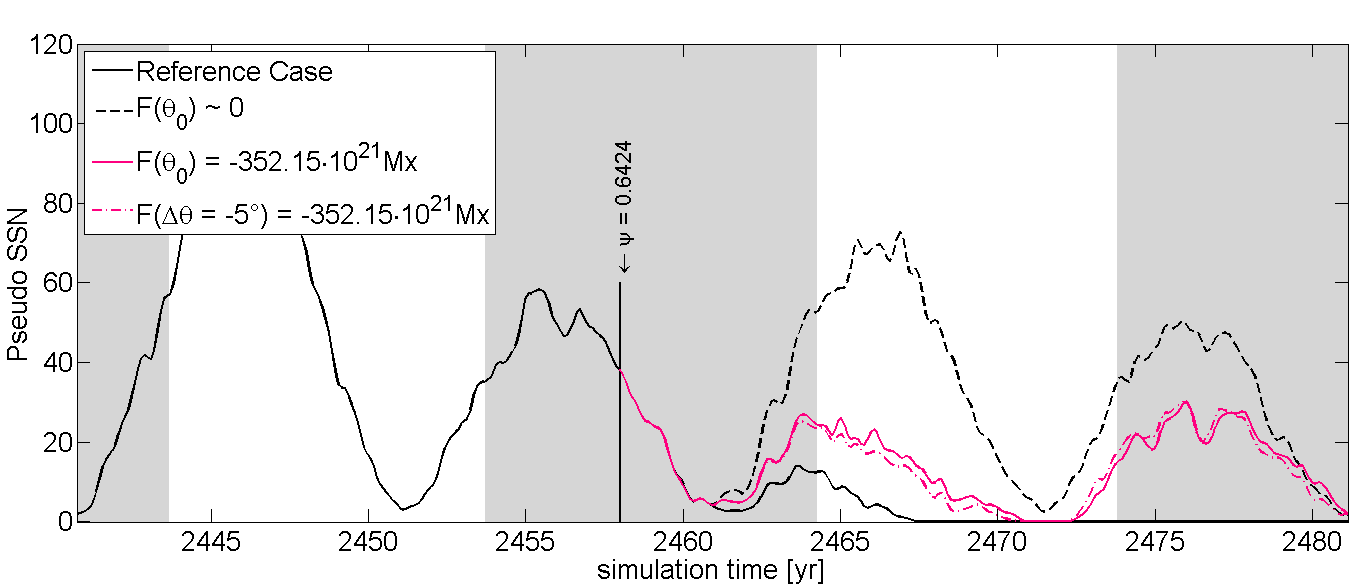}\\
      \includegraphics[width=\textwidth]{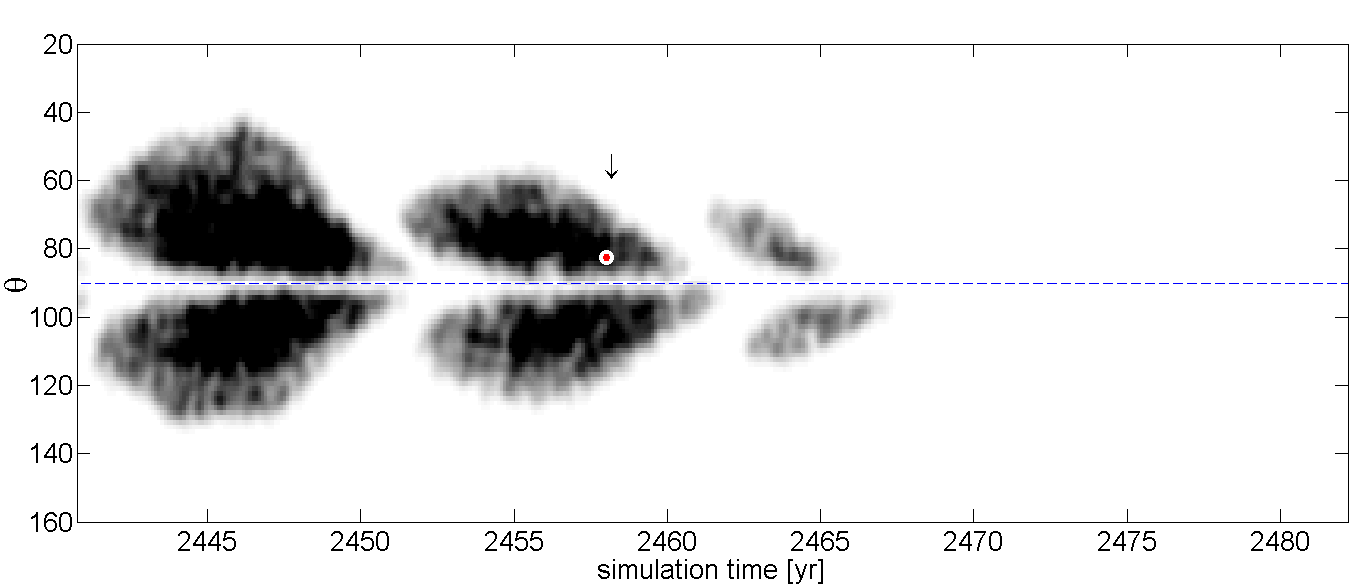}\\
      \caption{The spot that killed the dynamo: extreme effect of a huge
      active region during the descending phase of the cycle, on the
      SSN curve (top) and on the butterfly diagram (bottom). Red dot indicates the
      position of the leading polarity of the rogue BMR.
      The black dashed line in the top panel shows how the simulation
      unfolds if the rogue BMR is removed. The gray background
      corresponds to the negative dipole periods of this simulation as
      the dynamo in the reference case shuts down. The other curves indicate
      other simulations where the flux of the BMR was decreased (solid pink)
      and where this BMR with decreased flux emerged closer to the equator
      by 5$^{\circ}$. }
      \label{fig:shutdownDynamo}
    \end{figure}

In the case plotted in Figure \ref{fig:shutdownDynamo} the dynamo is
turned off by a BMR of extremely high flux and unfavorable tilt angle
emerging after cycle maximum (see column 3 of Table \ref{tab:BMRs}).
\RA{We denote the cycle phase of the emergence of the active region by
$\psi$, the cycle phase being defined as the ratio
of the time since the minimum and the duration of the cycle.}
This active region triggers an early polarity reversal which finally
occurs during the same pseudo-sunspot cycle. After this event, the
dipole moment does not build up again, instead
decaying to zero during
the following twenty years. The dashed black curve in Figure
\ref{fig:shutdownDynamo} shows that the dynamo keeps working if the
identified peculiar BMR is removed from the simulation.

Two further experiments were done in this case in order to see whether
or not a BMR with lower flux could cause the same effect at the same
position, or closer to the equator. As it can be seen in Figure
\ref{fig:shutdownDynamo}, an active region with a flux one half of the
original BMR cannot produce the same effect, even if it is closer to
the equator.

Such stopping events of the dynamo action were also found during the
experiments detailed in Section \ref{sect:BMReffects}. The dipole moment
typically decayed after the test BMR emergence during the next 10-20
years,
commensurate with the magnetic diffusion timescale for the
bulk of the convection zone.

\subsection{Restart of Dynamo Action}\label{sect:restartDynamo}
    \begin{figure}[h!]
      \centering
      \includegraphics[width=\textwidth]{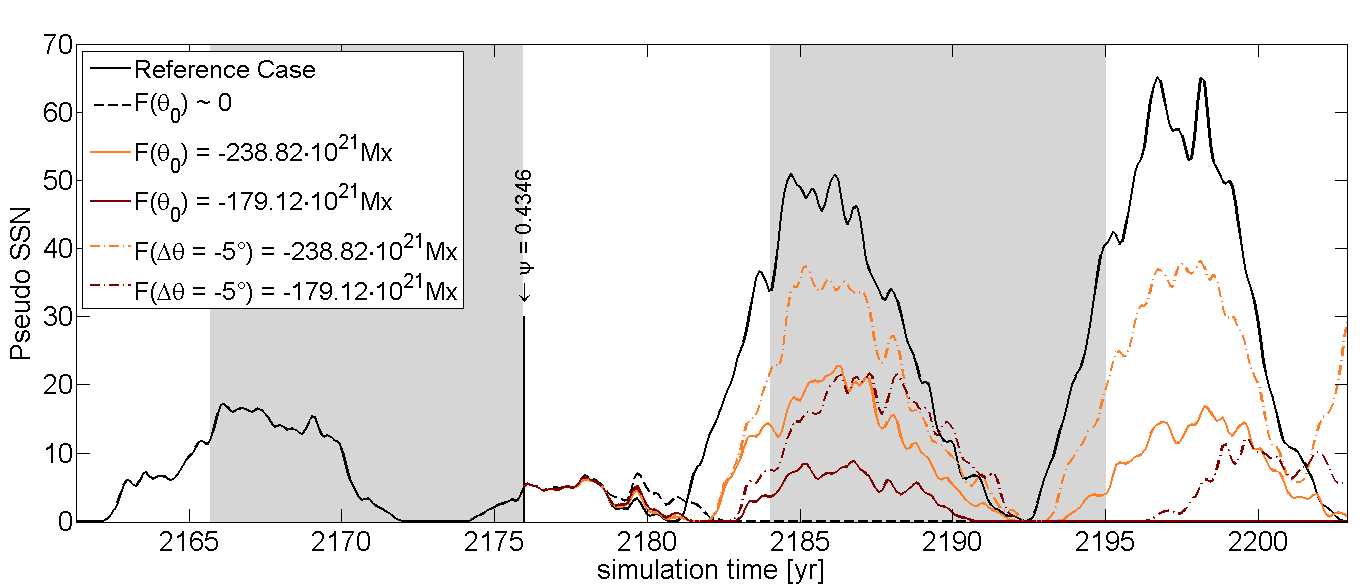}\\
      \includegraphics[width=\textwidth]{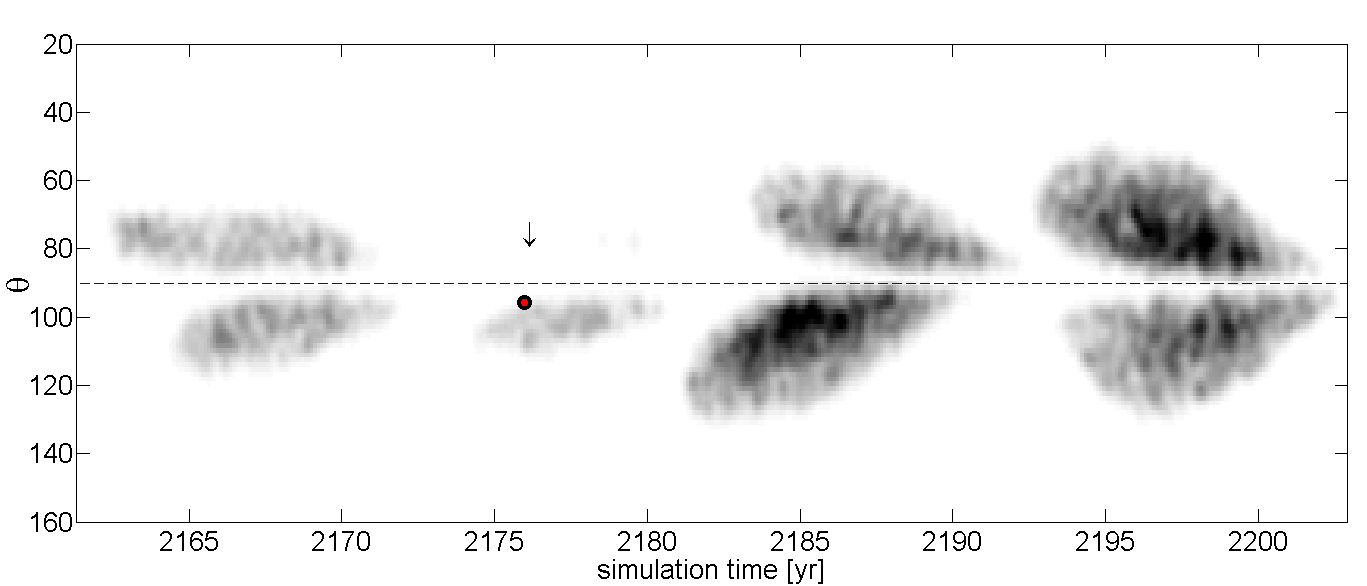}\\
      \caption{The spot that saved the dynamo: an example of the restart of
      an almost stopped dynamo simulation, due to large rogue BMR.
      Black dashed line shows
      the case when the active region was removed from the simulation. Solid
      curves of $F_{\mathrm{trail}}(\theta_0)$ indicate cases when the
      flux of the rogue BMR was reduced, while
      $F_{\mathrm{trail}}(\theta_0-5^{\circ})$ with dash-dot lines
      show the cases when the BMR was closer to the equator by 5$^{\circ}$.
      On the \emph{bottom} the butterfly diagram of the reference case is shown.
      Red dot indicates the position of the leading polarity of the
      rogue BMR.}
      \label{fig:restartDynamo}
    \end{figure}

We identified another extreme case, when the dying dynamo is
restarted by one peculiar BMR. In the reference case, indicated with
solid black line at $t = 2176$ in Figure \ref{fig:restartDynamo}, we
found an active region of high flux and favorable tilt angle
triggering an immediate polarity reversal and yielding a normal
amplitude dipole moment after that (see column 4 of Table
\ref{tab:BMRs} for the characteristics of this BMR). Without this
emergence the dipole moment would converge to zero in the next 20
years and no more active regions would emerge. This case is indicated
by a dashed black line in Figure \ref{fig:restartDynamo}.

We run a few more simulations in the case of this active region in
order to see whether or not a weaker BMR can produce the same result.
We first divided the flux value by 1.5 and 2 at the same position
(solid lines in Figure \ref{fig:restartDynamo}). At this position the
BMR with the smallest flux can yield to one more pseudo-SSN cycle, and
the dynamo action stops after this cycle. In the next step we inserted
the decreased flux BMRs closer to the equator (dot-dashed lines in
Figure \ref{fig:restartDynamo}). The emergences
closer to the equator had a stronger effect, so that the weaker BMR
could also restart the dynamo.

\subsection{Effects on Hemispheric Asymmetry}\label{sect:asymmetry}

\begin{figure}[h!]
      \centering
      \includegraphics[width=\textwidth]{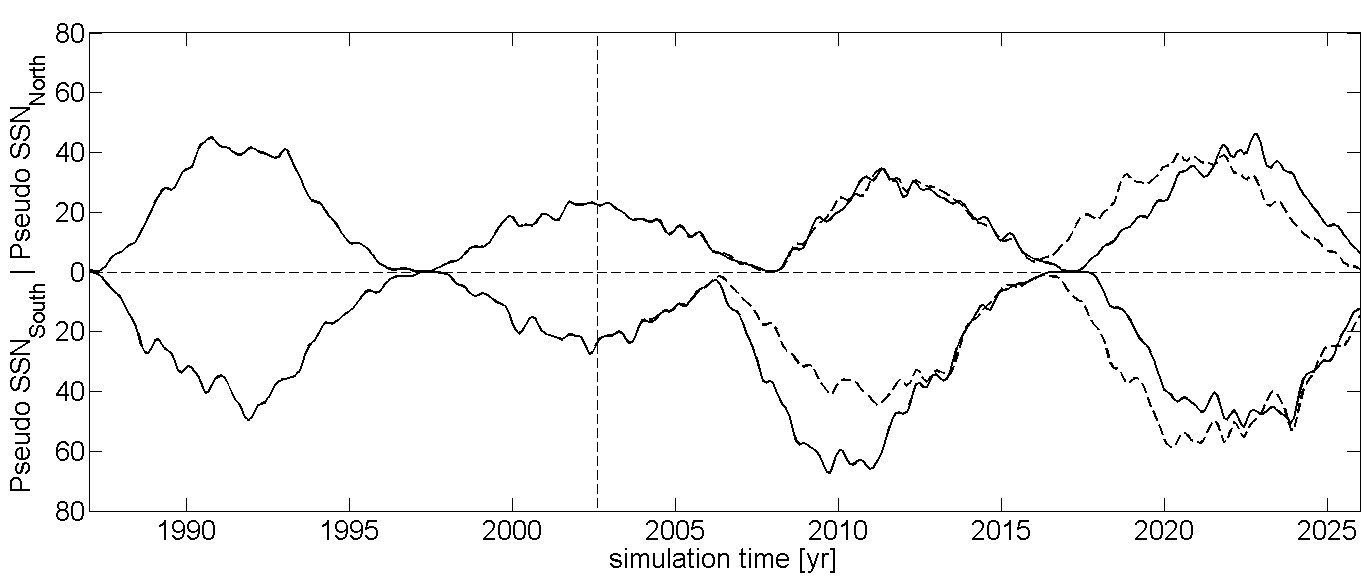}\\
      \includegraphics[width=\textwidth]{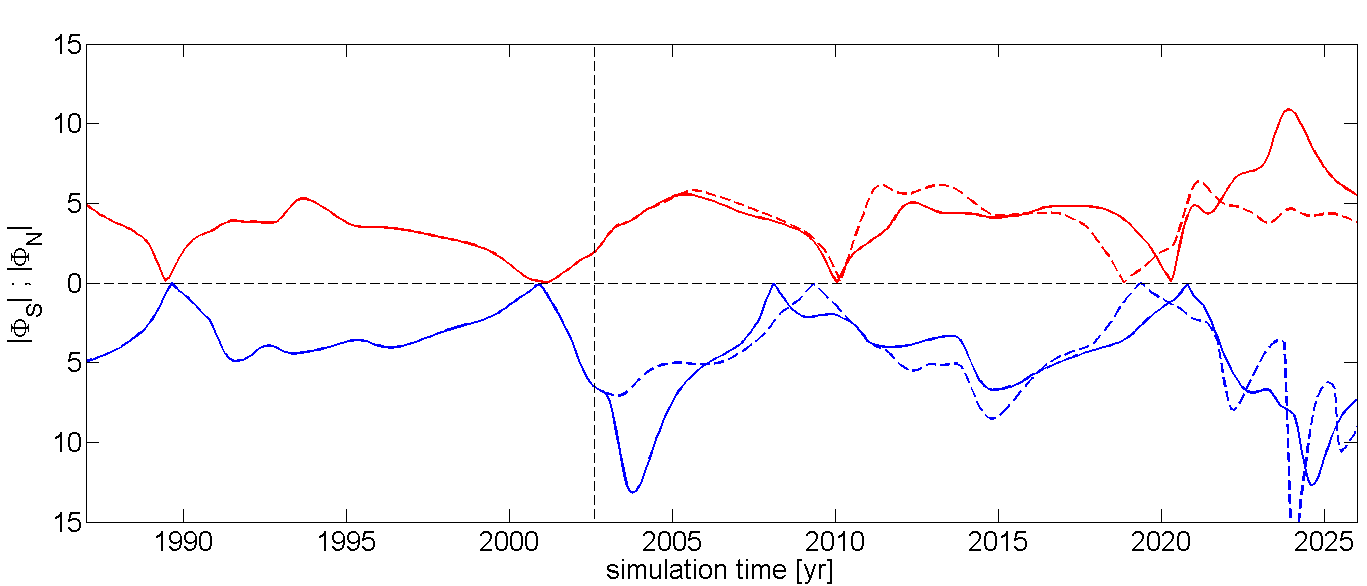}\\
      \caption{The \emph{top} figure shows an example pseudo-SSN time
series separately for the northern (+) and the southern hemispheres
(--) to illustrate to what extent a strong BMR affects the asymmetry
between the hemispheres. On the \emph{bottom} the absolute values of
the corresponding northern (+) (southern (--)) polar cap flux time
series is shown. In both panels the solid lines show the reference
case, while the dashed ones show the second run without the identified BMR.
The polar cap flux is calculated as the surface integral of the magnetic
field over a latitudinal extent of $20^{\circ}$ from the poles.
}\label{fig:asymmetry}
    \end{figure}

The butterfly diagrams of Figures \ref{fig:thenumber},
\ref{fig:shutdownDynamo} and \ref{fig:restartDynamo} suggest that
peculiar BMR emergences affect not only the amplitude or the starting
epoch of the subsequent cycle, but the hemispheric asymmetry as well.
In order to illustrate this effect, we chose the strongly asymmetric
cycle plotted with the solid line in the top panel of Figure
\ref{fig:asymmetry}. Besides the asymmetry in the activity levels of
the hemispheres we see in the third cycle, the new cycle starts about
three years earlier on the south. This asymmetry was caused by a rogue
BMR described in column 5 of Table \ref{tab:BMRs}. This active
region emerged relatively far from the equator during the maximum of
the second cycle, but still close enough to see significant effect. As
the dashed line in the top panel of Figure \ref{fig:asymmetry} shows,
by removing the identified strong BMR from the simulation, the
northern hemisphere in the third cycle shows the same amount of active
regions as before, but in the south one can see 30\% lower amplitude
compared to the reference case. According to \citet{Hathaway2016}, this
asymmetry we see in this cycle might be predicted by the polar cap
flux asymmetry during the perturbed cycle. As our plot shows in the
bottom panel of Figure \ref{fig:asymmetry}, the asymmetry in the
reference case already appeared in the second cycle in the form of
polar cap flux asymmetry. After the rogue BMR was removed, the polar
cap flux asymmetry decreased in the course of the second cycle and
consequently, the activity asymmetry was reduced in the third cycle as well.

To quantify this result we selected a series of 540 simulated cycles and compared the normalized asymmetry of the peak polar cap flux produced during the cycles ($\Delta_{\Phi}$) to two asymmetry parameters of the subsequent cycles. These parameters were the asymmetry of the total number of emergences at each hemisphere ($\Delta_{\mathrm{SSN}}$), and the time delay between the epochs when the new BMRs started to emerge on the North and the South ($\Delta_{T}$).

    The asymmetry of the polar cap flux at a given cycle is defined as follows:
        \begin{equation}
        \Delta_{\Phi} = \frac{|\Phi_{\mathrm{N},\mathrm{max}}| - |\Phi_{\mathrm{S},\mathrm{max}}|} {( |\Phi_{\mathrm{N},\mathrm{max}}| + |\Phi_{\mathrm{S},\mathrm{max}}| )/2},
        \end{equation}
    \noindent where $\Phi_{\mathrm{N},\mathrm{max}}$ ($\Phi_{\mathrm{S},\mathrm{max}}$) is the northern (Southern) polar cap flux maximum.

    The asymmetry of the activity level:
        \begin{equation}
        \Delta_{\mathrm{SSN}} = \frac{\Sigma \mathrm{SSN}_{\mathrm{N}} - \Sigma \mathrm{SSN}_{\mathrm{S}} } {( \Sigma \mathrm{SSN}_{\mathrm{N}} + \Sigma \mathrm{SSN}_{\mathrm{S}} )/2},
        \end{equation}
    \noindent where $\Sigma \mathrm{SSN}_{\mathrm{N}}$ ($\Sigma \mathrm{SSN}_{\mathrm{S}}$) is the total number of emergences on the northern (southern) hemisphere.

    The time lag between the hemispheres:
        \begin{equation}
          \Delta_{T} = \frac{t_{\mathrm{N}} - t_{\mathrm{S}}} {( T_{\mathrm{N}} + T_{\mathrm{S}} )/2},
        \end{equation}
    \noindent where $t_{\mathrm{N}}$ ($t_{\mathrm{S}}$) is the beginning epoch of the cycle, while $T_{\mathrm{N}}$ ($T_{\mathrm{S}}$) is the duration of the cycle on the North (South).

    \begin{figure}[h!]
      \centering
      \includegraphics[width=0.65\textwidth]{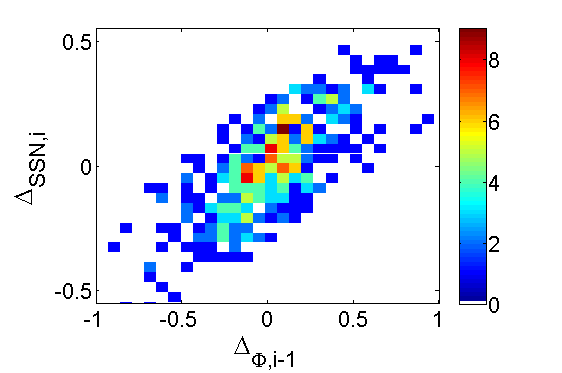}\\
      \includegraphics[width=0.65\textwidth]{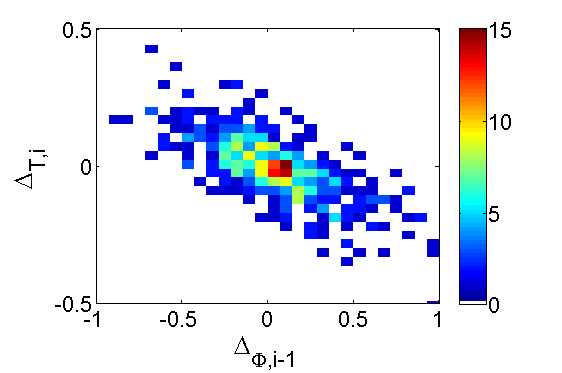}\\
      \caption{\RA{
      Two-dimensional histograms of the asymmetry of the hemispheric total
     pseudo-SSN
     (top) and the time lag between North and South (bottom) in
     pseudo-solar cycle $i$ against the polar cap flux asymmetry during the
     previous cycle. Some outlier data have been removed.
     The number of cases (cycles) in each bin are
     indicated by the colour codes. The correlation coefficients are
     $0.7430$ and $-0.7174$, respectively.}}
\label{fig:AsymmetryCrossplot}
    \end{figure}

    Figure \ref{fig:AsymmetryCrossplot} shows \RA{the results in the form of two-dimensional count histograms}.
The correlation between the polar cap flux asymmetry during cycle $i$
and the activity asymmetry during cycle $i+1$ is $r = 0.7430$. In the
case of the time lag obtained during cycle $i+1$ is $r = -0.7174$.
According to this result, the asymmetry level of the polar fields that
build up during the cycles is a good predictor of the asymmetry for
the subsequent cycle.

Hemispheric asymmetries were previously mostly discussed in the context of
variations in the meridional flow (\citealt{Belucz2013}, \citealt{Hathaway2016}).
The possibility that hemispheric asymmetries may be induced by rogue active
regions offers an interesting and viable alternative, to be explored in more
detail in future research.



\section{Effects of a Rogue Active Region: a Systematic Analysis}
\label{sect:BMReffects}

In previous sections we discussed cases of rogue BMRs that
spontaneously arose during the course of the simulation. In order to
get a more systematic assesment of the effect of rogue BMR, in this
section we will present results from a series of numerical
experiments where an additional, ``test'' BMR was manually inserted
into ongoing simulations with preset characteristics.  The properties
of this BMR are given in the sixth column of Table \ref{tab:BMRs}. (A
BMR with such characteristics did actually emerge  during the
reference simulation at an earlier epoch.)

The investigation was performed for three cycles of average, below
average and above average amplitudes, respectively. In each
case two series of experiments were
carried out with  Hale (anti-Hale) test BMR in order to increase (decrease) the dipole
moment of the examined cycle. As results were largely coherent among
these series, in the accompanying plots only one of these cases (a
test BMR which decreases the dipole moment of an average amplitude
cycle) is shown.

In each series of experiments one parameter of the test BMR was
systematically varied: emergence time, emergence latitude, magnetic
flux, angular separation, or tilt angle.


\subsection{Effects of Active Region Emergence Timing}
    \begin{figure}[h!]
      \centering
      \includegraphics[width=\textwidth]{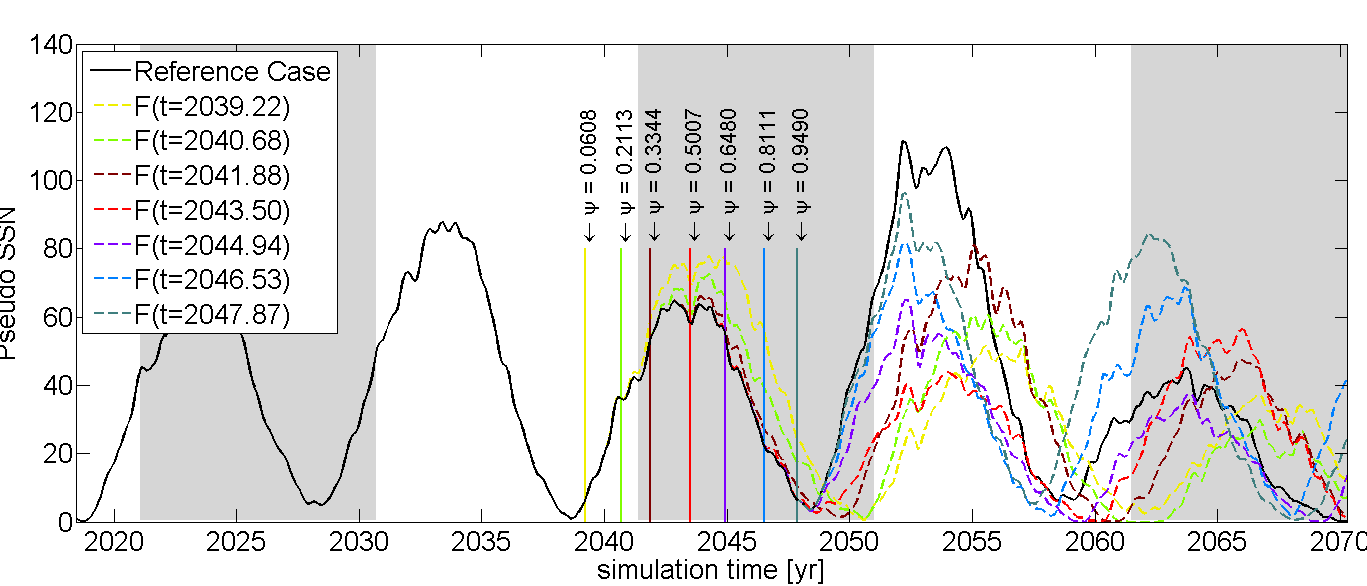}\\
      \includegraphics[width=\textwidth]{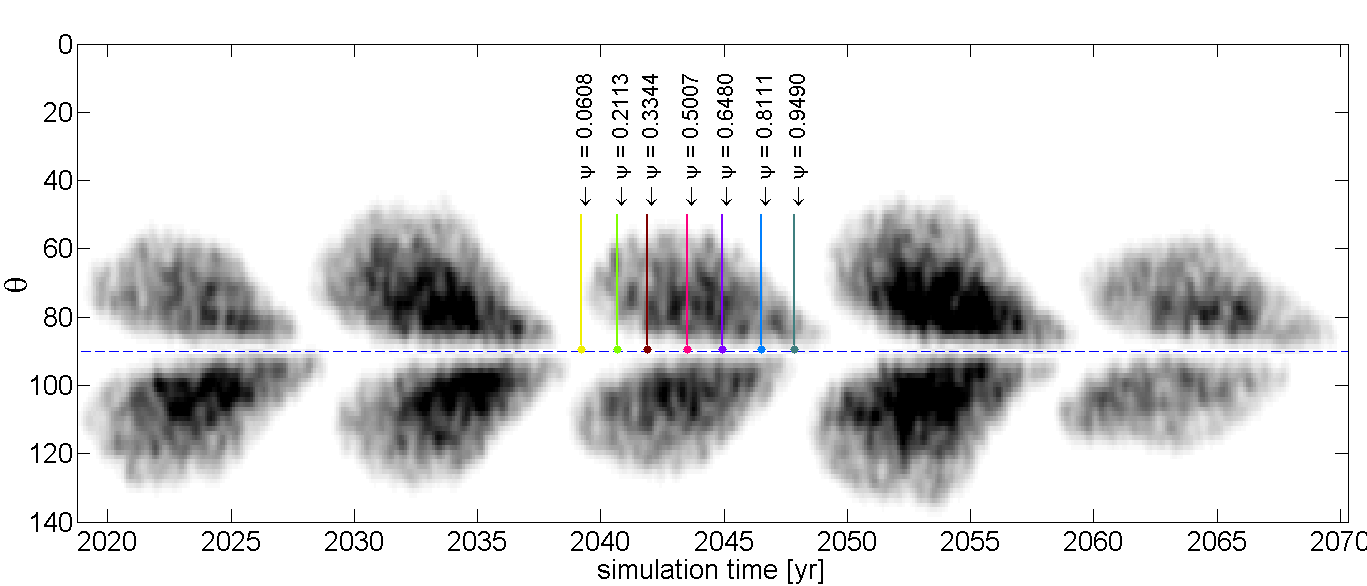}\\
      \caption{ The \emph{top} figure illustrates the effect of a single BMR inserted in the simulation at different phases of the cycle. The black solid line represents the reference case without the test BMR, gray background indicates the negative phase of the dipole moment. On the \emph{bottom} the butterfly diagram of the reference case is shown. The dots indicate the position of the leading polarity of the active region. The properties of the active region are listed in column 6 of Table \ref{tab:BMRs}.}
      \label{fig:experimentPsi}
    \end{figure}

During the first experimental series the emergence time of the test
BMR was varied. The emergence epochs were chosen based on the phase
of the cycle $\psi$. 
In the example case shown in
Figure \ref{fig:experimentPsi}, the inserted BMR decreases the dipole
moment during the course of an average magnitude cycle.

If a peculiar BMR emerges during the rising phase of the simulated
cycle it affects the amplitude of the ongoing cycle as well. The
explanation to this is that peculiar BMRs occurring before dipole-moment polarity-reversal can modify its timing. Obviously, after the
time of maximum the impact of such BMRs on the current cycle
disappears. Since during the rising phase these active regions affect
the ongoing cycle itself, their probable effect on the subsequent one
will be unpredictable.

If the emergence happens during the descending phase, it tends to have
less impact on the subsequent cycle. It is clear from Figure
\ref{fig:experimentPsi} that the later the BMR emerges, the smaller
the amplitude change expected during the next cycle. The strongest
effect is expected if the perturbation occurs at cycle maximum.  The
amplitude of the next cycle can be lower (higher) by up to 100\% when
the emergence decreases (increases) the amplitude of the dipole moment
being built up.

The length of the current cycle can also be modified but the impact
decreases during the rising phase and after the pseudo-SSN maximum it
disappears entirely. The ongoing cycle can be lengthened (shortened)
by up to two years if the emerged peculiar BMR decreases (increases) the
amplitude of the dipole moment. Due to this duration change, the
beginning of the subsequent cycle also shifts by up to two years. This
result is particularly interesting, since in this dynamo model the
meridional flow speed is constant, and sets the cycle length in a simulation in which stochasticity in properties of BMRs is turned off.
In general, not only the magnitude of the dipole moment
changes, but also the timing of polarity reversals.

Emergences of peculiar BMRs before and after the maxima also
result in different timings of the subsequent simulated cycle maximum
due to the the shifting of the minimum epochs.

\subsection{Effects of Active Region Flux}\label{sect:flux}
    \begin{figure}[h!]
      \centering
      \includegraphics[width=\textwidth]{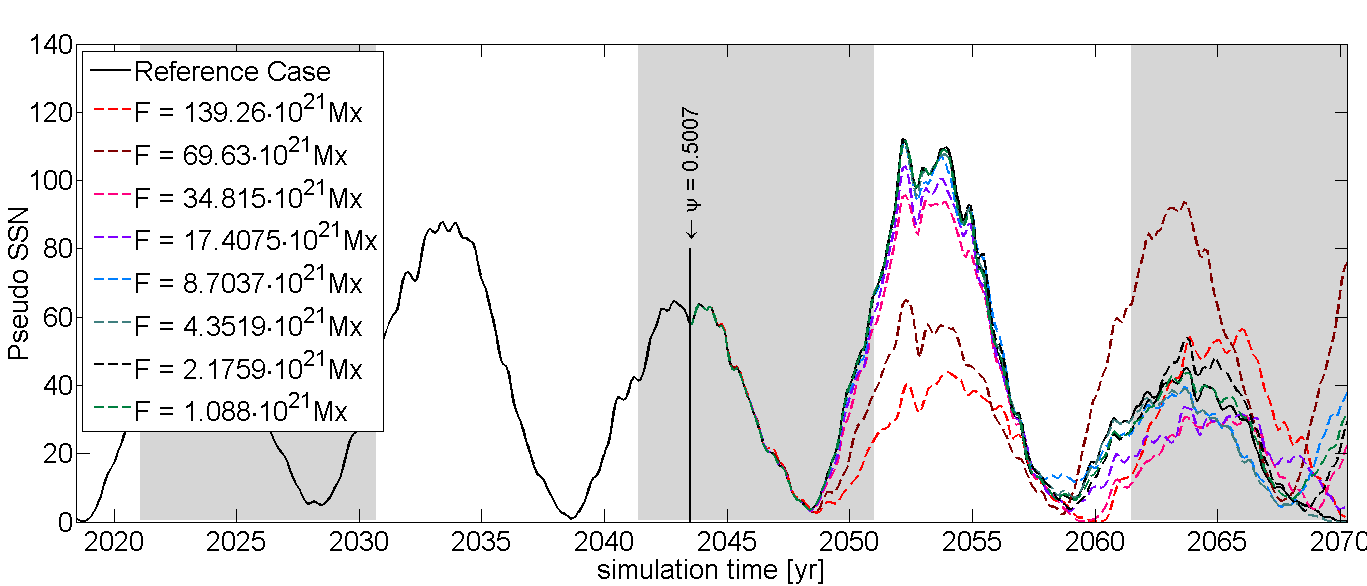}\\
      \caption{ The figure shows how the impact of a single inserted BMR at cycle maximum changes depending on its flux. The red curve, corresponding to the lowest amplitude during the subsequent cycle, is identical to the $\psi = 0.5007$ case on Figure \ref{fig:experimentPsi}. Gray background indicates the negative phase of the dipole moment. The properties of the active region are listed in column 6 of Table \ref{tab:BMRs}.}
      \label{fig:experimentDflux}
    \end{figure}

The second set of simulation runs focuses on the role of the magnetic flux
of the emerging active region. Here, the flux of the test BMR
is decreased in logarithmic increments (a factor of two beween
adjacent cases), while other
parameters are kept constant. The test BMR is inserted at cycle
maximum where the highest impact is expected. Figure
\ref{fig:experimentDflux} shows that the effect on the
amplitude of the subsequent cycle
decreases with decreasing magnetic flux of the test BMR.

On the other hand, this modification of the flux
does not cause significant changes to the minimum epoch. As
suggested in Figure \ref{fig:experimentDflux} above, the subsequent
cycle tends to start producing new active regions earlier (later) due
to the effect of inserted active regions increasing (decreasing) the
dipole moment. Identically, the duration of the minimum tends to be
shorter (longer), when the test BMR increased (decreased) the dipole
moment. The minimum duration is defined by the period when the pseudo-SSN is below 12.5. This is twice the lowest activity seen at
$t = 2058.5$.

\subsection{Effects of Tilt and Angular Separation}\label{sect:tiltsepa}
    \begin{figure}[h!]
      \centering
      \includegraphics[width=\textwidth]{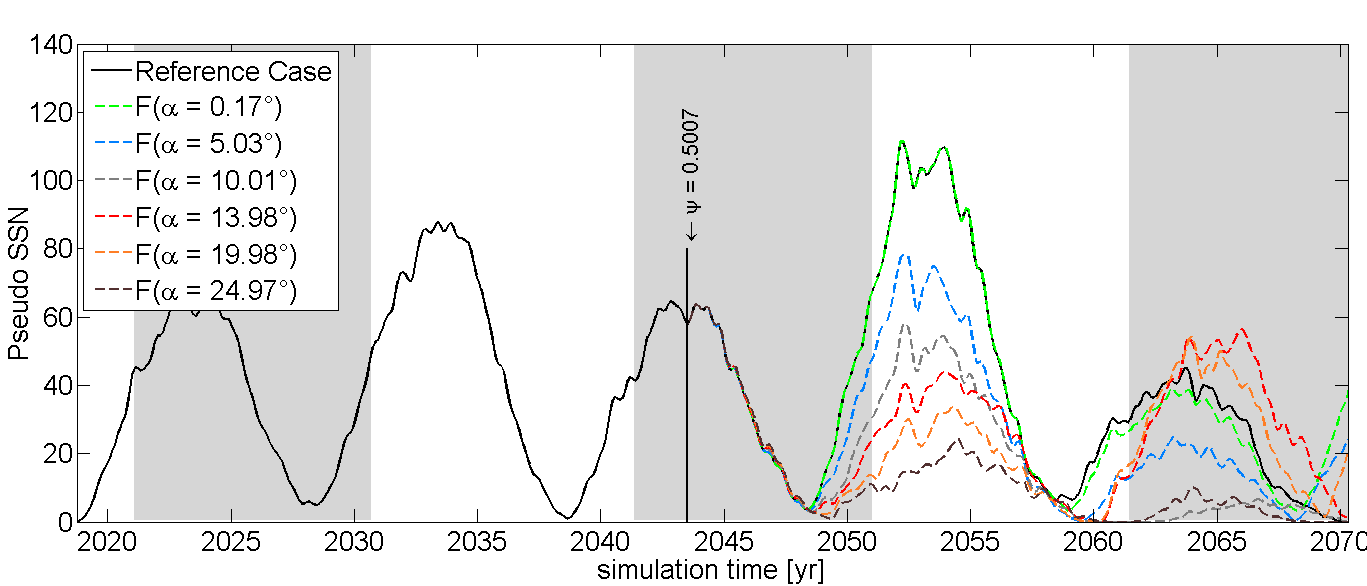}\\
      \caption{The plot shows how the tilt angle of the inserted BMR at cycle maximum affects the amplitude of the next cycle. The red curve is identical to the $\psi = 0.5007$ case on Figure \ref{fig:experimentPsi}. Gray background indicates the negative phase of the dipole moment. The properties of the active region are listed in column 6 of Table \ref{tab:BMRs}.}
      \label{fig:experimentDtilt}
    \end{figure}
The importance of the latitudinal size, \emph{i.e.} the tilt angle and the
separation of the inserted active region polarities were also
examined. As an example we plot a result from perturbing the tilt
angle in Figure \ref{fig:experimentDtilt}. As the
tilt angle of the active region was increased from $0^{\circ}$ to
$25^{\circ}$ in $\Delta\alpha = 5^{\circ}$ steps, the amplitude of the
subsequent cycle changed to a higher extent relative to the reference
case. Evidently, for $\alpha = 0^{\circ}$ the amplitude change
of the next cycle is zero as the net contribution to the dipole
moment is zero. For the case of the emerging test BMR with angular
separation and flux parameters listed in column 6 of Table
\ref{tab:BMRs}, at $\alpha = 25^{\circ}$ the amplitude of the
subsequent simulated cycle can increase by up to 200\% in cases when
the emerged BMR increased the dipole moment. When the emergence
decreases the dipole moment the dynamo action tends to stop in the case
of weak following cycles.

The separation of the active region was changed in steps of $\Delta d
= 5^{\circ}$ from $\approx 16^{\circ}$ to $\approx 41^{\circ}$. This kind of
perturbation yields the same result as changing the tilt angle, since
in both cases the latitudinal size of the active region changes, which
contributes to the efficiency of the polarity separation by the
meridional circulation.

Similarly to the case of flux perturbation, active regions of the next cycle tend to emerge earlier (later) when the test region increased (decreased) the dipole moment. This effect, however, is more prominent if the perturbed property is the tilt angle, as Figure \ref{fig:experimentDtilt} shows.

\subsection{Effects of Changes in Latitude}\label{sect:latitude}
    \begin{figure}[h!]
      \centering
      \includegraphics[width=\textwidth]{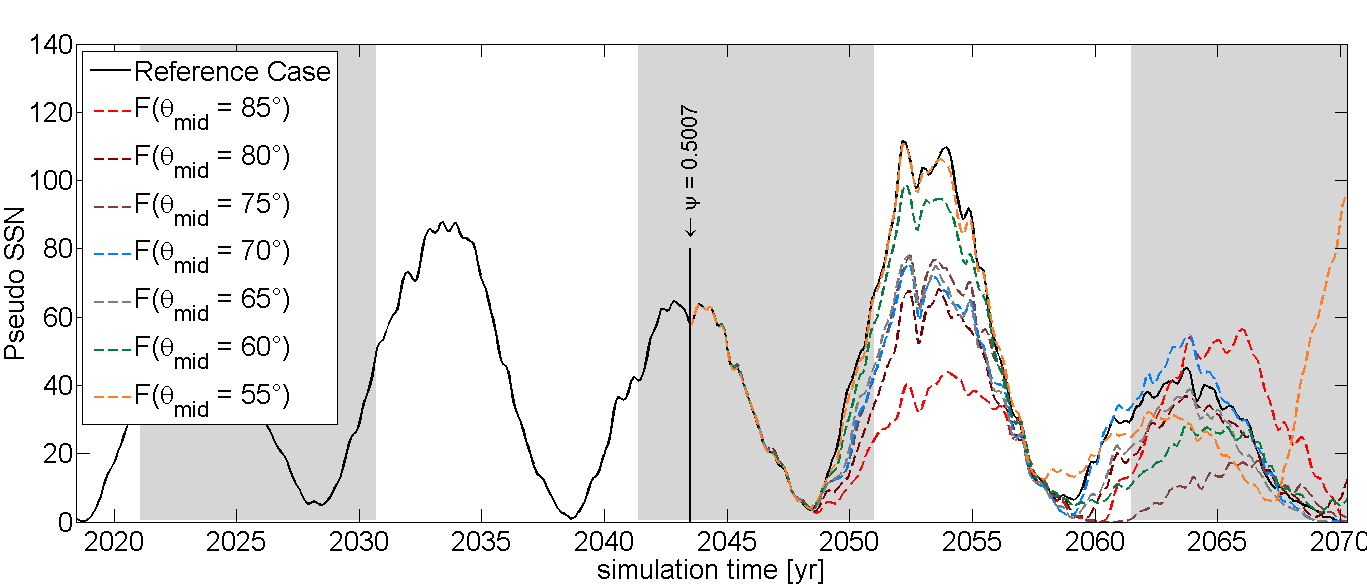}\\
      \caption{ The figure shows how the latitude of the inserted BMR at cycle maximum affects the amplitude of the subsequent cycle. The red curve, corresponding to the lowest amplitude during the subsequent cycle, is identical to the $\psi = 0.5007$ case on Figure \ref{fig:experimentPsi}. Gray background indicates the negative phase of the dipole moment. The properties of the active region are listed in column 6 of Table \ref{tab:BMRs}.}
      \label{fig:experimentDtheta}
    \end{figure}

According to \citet{Cameron2014} active regions very close to the
equator have significant effect on the dipole moment. To verify this
for our model we investigated the impact of decreasing the colatitude
of the leading and trailing polarities in steps of $|\Delta\theta| =
5^{\circ}$ from $\theta_{\mathrm{mid}} = 85^{\circ}$ to $\theta_{\mathrm{mid}} =
55^{\circ}$. We found that $20^{\circ}$ from the equator there is
still about a 50\% effect on the amplitude of the subsequent cycle.
Note that although the angular separation stays the same by
definition, the linear distance will decrease by 17\% from the
original to the highest position. The error caused by the decreased
physical size is only 5\% at $\theta_{\mathrm{mid}} = 70^{\circ}$, \emph{i.e.} $20^{\circ}$
 from the equator, where the effect on the next cycle is still
significant.

In terms of changes in the length of the next cycle no systematic
effect was found.


\subsection{Discussion}

\begin{figure}[h!]
      \centering
      \includegraphics[width=0.65\textwidth]{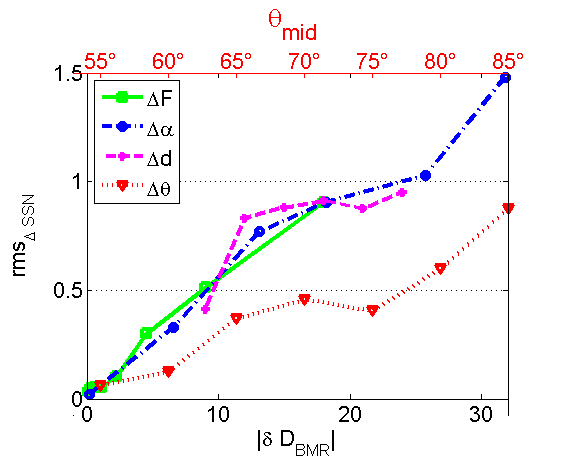}
      \includegraphics[width=0.65\textwidth]{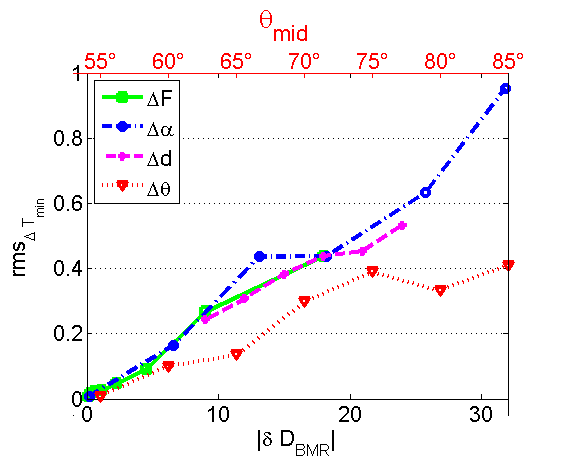}
      \caption{ Average effect of BMR emergences on the amplitude of the subsequent cycle (top panel) and on the duration of the minimum (bottom panel). The flux change (green curve), the tilt angle variation (blue curve) and the separation modification (magenta curve) of the active region are converted to the contribution to the build up of the dipole moment according to Equation \ref{eq:thenumber}. The top axis, corresponding to the red dotted curve, indicates the colatitude of the BMR. } \label{fig:summarise2}
    \end{figure}

\RA{
Substituting an individual BMR into Equation \ref{eq:dipolemoment},
one finds that, to leading order, the contribution of this source to
the global dipole moment scales as
\begin{equation}\label{eq:thenumber}
  \delta D_{\mathrm{BMR}} \approx F \,d \, \sin\alpha \,
  \sin\theta,
\label{eq:deltaD}
\end{equation}
where, as before, $d$ is the angular separation of leading and trailing
polarities, $\alpha$ is the tilt angle, \RA{$\theta$ is the co-latitude
of the active region midpoint,}
and $F=F_{\mathrm{trail}}$ is the flux of the trailing polarity
($F_{\mathrm{trail}}=-F_{\mathrm{lead}}$).

This suggests that the effects of the individual factors in Equation
\ref{eq:thenumber} can be combined in $\delta D_{\mathrm{BMR}}$. This
is indeed borne out in Figure \ref{fig:summarise2}. The top abscissa
of both panels shows the middle colatitude of the test BMR inserted
during the simulations discussed in Section \ref{sect:latitude}. The
bottom abscissa shows the contribution of the test active region to the
global dipole moment ($\delta D_{\mathrm{BMR}}$) converted by Equation
\ref{eq:thenumber} from the flux, tilt angle and separation changes
described in Sections \ref{sect:flux} and \ref{sect:tiltsepa}. In each
case the constant parameters were taken from the sixth column of Table
\ref{tab:BMRs}. The ordinates in Figure \ref{fig:summarise2} show the
relative change in amplitude of the next cycle (top panel) and the
duration of the
minimum (bottom panel) changes averaged for the simulation runs.
The new cycle properties after inserting the test BMR,
$X_{\mathrm{test}}$ (\emph{e.g.} amplitude) are always compared to the
reference case, $X_{\mathrm{ref}}$ as follows: $\mathrm{rms}_{\Delta}
= \sqrt{\left\langle \left( \frac{ X_{\mathrm{ref}}-X_{\mathrm{test}}
}{X_{\mathrm{ref}}} \right)^2 \right\rangle }$.

The good agreement of the green, magenta and blue curves
of Figure \ref{fig:summarise2} confirms that the contribution number
$\delta D_{\mathrm{BMR}}$ is indeed an appropriate combined measure of
the ``dynamo efficiency'' of individual active regions.
On the observational front,
the importance of the combined influence of $F$ and $\alpha$ on the
buildup of the poloidal field was demonstrated by \cite{Munoz2013}
based on mean-tilt-weighted active region areas
and polar facular counts as a proxy of the dipole moment for Cycles 15--23.
}

The red dotted curve showing the effect of emergence latitude clearly
indicates that the impact on the peak value of the subsequent cycle
decreases if the emergence occurs farther away from the equator.
Nevertheless, in the model analyzed here BMR emerging around
20$^\circ$ latitude can still have significant effect on the
subsequent cycle.

\RA{
In this context it should be emphasized that Equation
\ref{eq:deltaD} only yields the {\it initial} contribution of the
BMR to the dipole moment. As time proceeds, this contribution is
significantly modified by diffusion and advection on the solar
surface, until an asymptotic value is reached. This process was
studied by \citet{Jiang2014} in a surface flux transport model.
They find (their Figure 6) that the final contribution to the global
dipole decreases quite fast with increasing emergence latitude,
becoming negligible at 15--20$^\circ$. (See also
\citealt{Jiang2014SSRv}.) These findings are in qualitative agreement
with our results, although some quantitative differences, probably related to
model details, remain to be clarified.
}

On the other hand, \citet{Hazra2017} report the opposite trend, based on
a small set of numerical experiments similar to those described herein.
While the origin of this discrepancy remains unclear at this juncture,
it is worth pointing out that in their observational analysis
\citet{Yeates2015} find that high latitude BMRs do not have
a long-term contribution to the polar field, lending support to the
tendency reported in this paper and in \citet{Jiang2014}.

\section{Conclusion}

Using a new ``2$\times$2D'' dynamo model of the solar activity cycle
we studied the effect of rogue BMRs (large BMRs with unusual
characteristics) on subsequent simulated solar cycles. We found that
even a single rogue BMR can have a major effect on the further
development of solar activity cycles, boosting or suppressing the
amplitude of subsequent cycles. In extreme cases an individual BMR can
completely halt the dynamo, triggering a grand minimum. Alternatively a
dynamo on the verge of being halted can also be resuscitated by a
rogue BMR with favourable characteristics. Rogue BMRs also have the
potential to induce significant hemispheric asymmetries in the solar
cycle.

To study the effect of rogue BMRs in a more systematic manner, we have
conducted a series of dynamo simulations where a large test BMR is
manually introduced in the model at various phases of cycles of
different amplitudes. A BMR emerging in the rising phase of a cycle
can modify the amplitude of the ongoing cycle while a BMR emerging in
later phases only impact subsequent cycles. The strongest
impact on the subsequent cycle is found when the rogue BMR emerges
around cycle maximum at low latitudes but the BMR does not need to be
strictly cross-equatorial.  Active regions emerging as far as
$20^{\circ}$  from the equator still have a significant effect on the
subsequent cycle. We have demonstrated that the combined effect of the
magnetic flux, tilt angle and polarity separation of the BMR on the
dynamo is via the contribution to the dipole moment, $\delta D_{\mathrm{BMR}}$.
An increase in either of these quantities will lead to an enhancement
of the effect on solar activity in the subsequent cycle. The sense of the
effect, in turn, depends on the sign of the contribution to the dipole
moment being built up in the late phases of the cycle.

The selected cases discussed in Sections 3 and 4 are in agreement with
the general rules derived in Section 5.  For instance, in panel (C) of
Figure~\ref{fig:thenumber}, at the maximum of the second cycle in the
reference simulation the identified active region had a decreasing
effect on the dipole moment, hence the magnitude of the subsequent
pseudo-SSN cycle also decreased. In the second run, indicated by
dashed black line, the identified BMR was removed from the simulation.
As a consequence, the third cycle doubled in amplitude compared
to the reference case. Besides this, the duration of
the minimum decreased significantly, as well. In terms of hemispheric
asymmetry, as it is shown in panel (B) for the reference case, the BMR
emerging on the north decreased the northern polar cap flux delaying the
onset of the subsequent pseudo-SSN cycle in the northern hemisphere.

It should be emphasized that while the rogue BMRs identified here are
large, they do not exceed in size the largest solar active regions on
record. Furthermore, as our detailed record only goes back to about a
score of solar cycles, the largest ever observed active regions by no
means represent a theoretical upper limit. Therefore, our results
suggest that large active regions with peculiar characteristics may be
responsible for unexpected changes in the behaviour of the solar
dynamo, as already suggested by \citet{Jiang2015}. More detailed
comparisons between simulated and observed solar cycles may be the
subject of future research.

 \begin{acks}
This project was partially funded by the European Union's Horizon 2020
research and innovation programme under grant agreement No. 739500,
by the Discovery Grant Program of the Natural Sciences and Engineering
Research Council of Canada, and by the Campus Mundi Program.

 \end{acks}

\noindent {\footnotesize \textbf{Conflict of interest} \indent
The authors declare that they have no conflicts of interest.}

 \bibliographystyle{spr-mp-sola}
 \bibliography{References}

\begin{thebibliography}{38}
\ifx\bisbn     \undefined \def\bisbn  #1{ISBN #1}\fi
\ifx\binits    \undefined \def\binits#1{#1}\fi
\ifx\bauthor   \undefined \def\bauthor#1{#1}\fi
\ifx\batitle   \undefined \def\batitle#1{#1}\fi
\ifx\bjtitle   \undefined \def\bjtitle#1{\textit{#1}}\fi
\ifx\bvolume   \undefined \def\bvolume#1{\textbf{#1}}\fi
\ifx\byear     \undefined \def\byear#1{#1}\fi
\ifx\bissue    \undefined \def\bissue#1{#1}\fi
\ifx\bfpage    \undefined \def\bfpage#1{#1}\fi
\ifx\blpage    \undefined \def\blpage #1{#1}\fi
\ifx\burl      \undefined \def\burl#1{\textsf{#1}}\fi
\ifx\href      \undefined \def\href#1#2{\textsf{#2}}\fi
\ifx\betal     \undefined \def\betal{\textit{et al.}}\fi
\ifx\bctitle   \undefined \def\bctitle#1{#1}\fi
\ifx\beditor   \undefined \def\beditor#1{#1}\fi
\ifx\bbtitle   \undefined \def\bbtitle#1{\textit{#1}}\fi
\ifx\bedition  \undefined \def\bedition#1{#1}\fi
\ifx\bseriesno \undefined \def\bseriesno#1{\textbf{#1}}\fi
\ifx\blocation \undefined \def\blocation#1{#1}\fi
\ifx\bsertitle \undefined \def\bsertitle#1{\textit{#1}}\fi
\ifx\bsnm      \undefined \def\bsnm#1{#1}\fi
\ifx\bsuffix   \undefined \def\bsuffix#1{#1}\fi
\ifx\bparticle \undefined \def\bparticle#1{#1}\fi
\ifx\barticle  \undefined \def\barticle#1{}\fi
\ifx\binstitute  \undefined \def\binstitute#1{#1}\fi
\ifx\bpublisher  \undefined \def\bpublisher#1{#1}\fi
\ifx\doiurl    \undefined
  \def\doiurl#1{\href{http://dx.doi.org/#1}{\textsf{DOI}}}\fi
\ifx\arxivurl  \undefined
  \def\arxivurl#1{\href{http://arxiv.org/abs/#1}{\textsf{arXiv}}}\fi
\ifx\adsurl    \undefined
  \def\adsurl#1{\href{http://adsabs.harvard.edu/abs/#1}{\textsf{ADS}}}\fi
\ifx\botherref \undefined \def\botherref#1{}\fi
\ifx\url       \undefined \def\url#1{\textsf{#1}}\fi
\ifx\bchapter  \undefined \def\bchapter#1{}\fi
\ifx\bbook     \undefined \def\bbook#1{}\fi
\ifx\bcomment  \undefined \def\bcomment#1{#1}\fi
\ifx\oauthor   \undefined \def\oauthor#1{#1}\fi
\ifx\citeauthoryear \undefined\def \citeauthoryear#1{#1}\fi
\ifx\endbibitem\undefined \def\endbibitem{}\fi
\ifx\bconflocation  \undefined \def\bconflocation#1{#1} \fi

\bibitem[\protect\citeauthoryear{{Baumann}, {Schmitt}, and
  {Sch{\"u}ssler}}{2006}]{Baumann2006}
\begin{barticle}
\bauthor{\bsnm{{Baumann}}, \binits{I.}},
\bauthor{\bsnm{{Schmitt}}, \binits{D.}},
\bauthor{\bsnm{{Sch{\"u}ssler}}, \binits{M.}}:
\byear{2006},
\batitle{{A necessary extension of the surface flux transport model}}.
\bjtitle{\aap}
\bvolume{446},
\bfpage{307}.
\doiurl{10.1051/0004-6361:20053488}.
\end{barticle}
\endbibitem

\bibitem[\protect\citeauthoryear{{Baumann} \textit{et~al.}}{2004}]{Baumann2004}
\begin{barticle}
\bauthor{\bsnm{{Baumann}}, \binits{I.}},
\bauthor{\bsnm{{Schmitt}}, \binits{D.}},
\bauthor{\bsnm{{Sch{\"u}ssler}}, \binits{M.}},
\bauthor{\bsnm{{Solanki}}, \binits{S.K.}}:
\byear{2004},
\batitle{{Evolution of the large-scale magnetic field on the solar surface: A
  parameter study}}.
\bjtitle{\aap}
\bvolume{426},
\bfpage{1075}.
\doiurl{10.1051/0004-6361:20048024}.
\end{barticle}
\endbibitem

\bibitem[\protect\citeauthoryear{{Belucz} and {Dikpati}}{2013}]{Belucz2013}
\begin{barticle}
\bauthor{\bsnm{{Belucz}}, \binits{B.}},
\bauthor{\bsnm{{Dikpati}}, \binits{M.}}:
\byear{2013},
\batitle{{Role of Asymmetric Meridional Circulation in Producing North-South
  Asymmetry in a Solar Cycle Dynamo Model}}.
\bjtitle{\apj}
\bvolume{779},
\bfpage{4}.
\doiurl{10.1088/0004-637X/779/1/4}.
\adsurl{http://esoads.eso.org/abs/2013ApJ...779....4B}.
\end{barticle}
\endbibitem

\bibitem[\protect\citeauthoryear{{Cameron} and
  {Sch{\"u}ssler}}{2015}]{Cameron2015}
\begin{barticle}
\bauthor{\bsnm{{Cameron}}, \binits{R.}},
\bauthor{\bsnm{{Sch{\"u}ssler}}, \binits{M.}}:
\byear{2015},
\batitle{{The crucial role of surface magnetic fields for the solar dynamo}}.
\bjtitle{Science}
\bvolume{347},
\bfpage{1333}.
\doiurl{10.1126/science.1261470}.
\end{barticle}
\endbibitem

\bibitem[\protect\citeauthoryear{{Cameron} \textit{et~al.}}{2010}]{Cameron2010}
\begin{barticle}
\bauthor{\bsnm{{Cameron}}, \binits{R.H.}},
\bauthor{\bsnm{{Jiang}}, \binits{J.}},
\bauthor{\bsnm{{Schmitt}}, \binits{D.}},
\bauthor{\bsnm{{Sch{\"u}ssler}}, \binits{M.}}:
\byear{2010},
\batitle{{Surface Flux Transport Modeling for Solar Cycles 15-21: Effects of
  Cycle-Dependent Tilt Angles of Sunspot Groups}}.
\bjtitle{\apj}
\bvolume{719},
\bfpage{264}.
\doiurl{10.1088/0004-637X/719/1/264}.
\end{barticle}
\endbibitem

\bibitem[\protect\citeauthoryear{{Cameron} \textit{et~al.}}{2014}]{Cameron2014}
\begin{barticle}
\bauthor{\bsnm{{Cameron}}, \binits{R.H.}},
\bauthor{\bsnm{{Jiang}}, \binits{J.}},
\bauthor{\bsnm{{Sch{\"u}ssler}}, \binits{M.}},
\bauthor{\bsnm{{Gizon}}, \binits{L.}}:
\byear{2014},
\batitle{{Physical causes of solar cycle amplitude variability}}.
\bjtitle{\jgr (Space Physics)}
\bvolume{119},
\bfpage{680}.
\doiurl{10.1002/2013JA019498}.
\end{barticle}
\endbibitem

\bibitem[\protect\citeauthoryear{{Charbonneau}, {St-Jean}, and
  {Zacharias}}{2005}]{Charbonneau2005}
\begin{barticle}
\bauthor{\bsnm{{Charbonneau}}, \binits{P.}},
\bauthor{\bsnm{{St-Jean}}, \binits{C.}},
\bauthor{\bsnm{{Zacharias}}, \binits{P.}}:
\byear{2005},
\batitle{{Fluctuations in Babcock-Leighton Dynamos. I. Period Doubling and
  Transition to Chaos}}.
\bjtitle{\apj}
\bvolume{619},
\bfpage{613}.
\doiurl{10.1086/426385}.
\end{barticle}
\endbibitem

\bibitem[\protect\citeauthoryear{{Charbonneau}
  \textit{et~al.}}{1999}]{Charbonneau1999}
\begin{barticle}
\bauthor{\bsnm{{Charbonneau}}, \binits{P.}},
\bauthor{\bsnm{{Christensen-Dalsgaard}}, \binits{J.}},
\bauthor{\bsnm{{Henning}}, \binits{R.}},
\bauthor{\bsnm{{Larsen}}, \binits{R.M.}},
\bauthor{\bsnm{{Schou}}, \binits{J.}},
\bauthor{\bsnm{{Thompson}}, \binits{M.J.}},
\bauthor{\bsnm{{Tomczyk}}, \binits{S.}}:
\byear{1999},
\batitle{{Helioseismic Constraints on the Structure of the Solar Tachocline}}.
\bjtitle{\apj}
\bvolume{527},
\bfpage{445}.
\doiurl{10.1086/308050}.
\end{barticle}
\endbibitem

\bibitem[\protect\citeauthoryear{{Choudhuri}, {Chatterjee}, and
  {Jiang}}{2007}]{Choudhuri2007}
\begin{barticle}
\bauthor{\bsnm{{Choudhuri}}, \binits{A.R.}},
\bauthor{\bsnm{{Chatterjee}}, \binits{P.}},
\bauthor{\bsnm{{Jiang}}, \binits{J.}}:
\byear{2007},
\batitle{{Predicting Solar Cycle 24 With a Solar Dynamo Model}}.
\bjtitle{\phrvl}
\bvolume{98},
\bfpage{131103}.
\doiurl{10.1103/PhysRevLett.98.131103}.
\end{barticle}
\endbibitem

\bibitem[\protect\citeauthoryear{{Dasi-Espuig}
  \textit{et~al.}}{2010}]{Dasi2010}
\begin{barticle}
\bauthor{\bsnm{{Dasi-Espuig}}, \binits{M.}},
\bauthor{\bsnm{{Solanki}}, \binits{S.K.}},
\bauthor{\bsnm{{Krivova}}, \binits{N.A.}},
\bauthor{\bsnm{{Cameron}}, \binits{R.}},
\bauthor{\bsnm{{Pe{\~n}uela}}, \binits{T.}}:
\byear{2010},
\batitle{{Sunspot group tilt angles and the strength of the solar cycle}}.
\bjtitle{\aap}
\bvolume{518}.
\doiurl{10.1051/0004-6361/201014301}.
\end{barticle}
\endbibitem

\bibitem[\protect\citeauthoryear{{Dikpati} and
  {Charbonneau}}{1999}]{Dikpati1999}
\begin{barticle}
\bauthor{\bsnm{{Dikpati}}, \binits{M.}},
\bauthor{\bsnm{{Charbonneau}}, \binits{P.}}:
\byear{1999},
\batitle{{A Babcock-Leighton Flux Transport Dynamo with Solar-like Differential
  Rotation}}.
\bjtitle{\apj}
\bvolume{518},
\bfpage{508}.
\doiurl{10.1086/307269}.
\end{barticle}
\endbibitem

\bibitem[\protect\citeauthoryear{{Dikpati} \textit{et~al.}}{2010}]{Dikpati2010}
\begin{barticle}
\bauthor{\bsnm{{Dikpati}}, \binits{M.}},
\bauthor{\bsnm{{Gilman}}, \binits{P.A.}},
\bauthor{\bsnm{{de Toma}}, \binits{G.}},
\bauthor{\bsnm{{Ulrich}}, \binits{R.K.}}:
\byear{2010},
\batitle{{Impact of changes in the Sun's conveyor-belt on recent solar
  cycles}}.
\bjtitle{\grl}
\bvolume{37},
\bfpage{L14107}.
\doiurl{10.1029/2010GL044143}.
\adsurl{http://esoads.eso.org/abs/2010GeoRL..3714107D}.
\end{barticle}
\endbibitem

\bibitem[\protect\citeauthoryear{{Fan}}{2009}]{Fan2009}
\begin{barticle}
\bauthor{\bsnm{{Fan}}, \binits{Y.}}:
\byear{2009},
\batitle{{Magnetic Fields in the Solar Convection Zone}}.
\bjtitle{\lrsp}
\bvolume{6}.
\doiurl{10.12942/lrsp-2009-4}.
\end{barticle}
\endbibitem

\bibitem[\protect\citeauthoryear{{Hathaway} and {Upton}}{2014}]{Hathaway2014}
\begin{barticle}
\bauthor{\bsnm{{Hathaway}}, \binits{D.H.}},
\bauthor{\bsnm{{Upton}}, \binits{L.}}:
\byear{2014},
\batitle{{The solar meridional circulation and sunspot cycle variability}}.
\bjtitle{Journal of Geophysical Research (Space Physics)}
\bvolume{119},
\bfpage{3316}.
\doiurl{10.1002/2013JA019432}.
\adsurl{http://esoads.eso.org/abs/2014JGRA..119.3316H}.
\end{barticle}
\endbibitem

\bibitem[\protect\citeauthoryear{{Hathaway} and {Upton}}{2016}]{Hathaway2016}
\begin{barticle}
\bauthor{\bsnm{{Hathaway}}, \binits{D.H.}},
\bauthor{\bsnm{{Upton}}, \binits{L.A.}}:
\byear{2016},
\batitle{{Predicting the amplitude and hemispheric asymmetry of solar cycle 25
  with surface flux transport}}.
\bjtitle{Journal of Geophysical Research (Space Physics)}
\bvolume{121},
\bfpage{10}.
\doiurl{10.1002/2016JA023190}.
\adsurl{http://esoads.eso.org/abs/2016JGRA..12110744H}.
\end{barticle}
\endbibitem

\bibitem[\protect\citeauthoryear{{Hazra}, {Choudhuri}, and
  {Miesch}}{2017}]{Hazra2017}
\begin{barticle}
\bauthor{\bsnm{{Hazra}}, \binits{G.}},
\bauthor{\bsnm{{Choudhuri}}, \binits{A.R.}},
\bauthor{\bsnm{{Miesch}}, \binits{M.S.}}:
\byear{2017},
\batitle{{A Theoretical Study of the Build-up of the Sun's Polar Magnetic Field
  by using a 3D Kinematic Dynamo Model}}.
\bjtitle{\apj}
\bvolume{835},
\bfpage{39}.
\doiurl{10.3847/1538-4357/835/1/39}.
\end{barticle}
\endbibitem

\bibitem[\protect\citeauthoryear{{Jiang}, {Cameron}, and
  {Sch{\"u}ssler}}{2014}]{Jiang2014}
\begin{barticle}
\bauthor{\bsnm{{Jiang}}, \binits{J.}},
\bauthor{\bsnm{{Cameron}}, \binits{R.H.}},
\bauthor{\bsnm{{Sch{\"u}ssler}}, \binits{M.}}:
\byear{2014},
\batitle{{Effects of the Scatter in Sunspot Group Tilt Angles on the
  Large-scale Magnetic Field at the Solar Surface}}.
\bjtitle{\apj}
\bvolume{791},
\bfpage{5}.
\doiurl{10.1088/0004-637X/791/1/5}.
\end{barticle}
\endbibitem

\bibitem[\protect\citeauthoryear{{Jiang}, {Cameron}, and
  {Sch{\"u}ssler}}{2015}]{Jiang2015}
\begin{barticle}
\bauthor{\bsnm{{Jiang}}, \binits{J.}},
\bauthor{\bsnm{{Cameron}}, \binits{R.H.}},
\bauthor{\bsnm{{Sch{\"u}ssler}}, \binits{M.}}:
\byear{2015},
\batitle{{The Cause of the Weak Solar Cycle 24}}.
\bjtitle{\apjl}
\bvolume{808},
\bfpage{L28}.
\doiurl{10.1088/2041-8205/808/1/L28}.
\adsurl{http://esoads.eso.org/abs/2015ApJ...808L..28J}.
\end{barticle}
\endbibitem

\bibitem[\protect\citeauthoryear{{Jiang} \textit{et~al.}}{2010}]{Jiang2010}
\begin{barticle}
\bauthor{\bsnm{{Jiang}}, \binits{J.}},
\bauthor{\bsnm{{I{\c s}ik}}, \binits{E.}},
\bauthor{\bsnm{{Cameron}}, \binits{R.H.}},
\bauthor{\bsnm{{Schmitt}}, \binits{D.}},
\bauthor{\bsnm{{Sch{\"u}ssler}}, \binits{M.}}:
\byear{2010},
\batitle{{The Effect of Activity-related Meridional Flow Modulation on the
  Strength of the Solar Polar Magnetic Field}}.
\bjtitle{\apj}
\bvolume{717},
\bfpage{597}.
\doiurl{10.1088/0004-637X/717/1/597}.
\end{barticle}
\endbibitem

\bibitem[\protect\citeauthoryear{{Jiang} \textit{et~al.}}{2014}]{Jiang2014SSRv}
\begin{barticle}
\bauthor{\bsnm{{Jiang}}, \binits{J.}},
\bauthor{\bsnm{{Hathaway}}, \binits{D.H.}},
\bauthor{\bsnm{{Cameron}}, \binits{R.H.}},
\bauthor{\bsnm{{Solanki}}, \binits{S.K.}},
\bauthor{\bsnm{{Gizon}}, \binits{L.}},
\bauthor{\bsnm{{Upton}}, \binits{L.}}:
\byear{2014},
\batitle{{Magnetic Flux Transport at the Solar Surface}}.
\bjtitle{\ssr}
\bvolume{186},
\bfpage{491}.
\doiurl{10.1007/s11214-014-0083-1}.
\end{barticle}
\endbibitem

\bibitem[\protect\citeauthoryear{{Lemerle} and
  {Charbonneau}}{2017}]{Lemerle2017}
\begin{barticle}
\bauthor{\bsnm{{Lemerle}}, \binits{A.}},
\bauthor{\bsnm{{Charbonneau}}, \binits{P.}}:
\byear{2017},
\batitle{{A Coupled 2$\times$2D Babcock-Leighton Solar Dynamo Model. II.
  Reference Dynamo Solutions}}.
\bjtitle{\apj}
\bvolume{834},
\bfpage{133}.
\doiurl{10.3847/1538-4357/834/2/133}.
\end{barticle}
\endbibitem

\bibitem[\protect\citeauthoryear{{Lemerle}, {Charbonneau}, and
  {Carignan-Dugas}}{2015}]{Lemerle2015}
\begin{barticle}
\bauthor{\bsnm{{Lemerle}}, \binits{A.}},
\bauthor{\bsnm{{Charbonneau}}, \binits{P.}},
\bauthor{\bsnm{{Carignan-Dugas}}, \binits{A.}}:
\byear{2015},
\batitle{{A Coupled $2\times2$D Babcock-Leighton Solar Dynamo Model. I. Surface
  Magnetic Flux Evolution}}.
\bjtitle{\apj}
\bvolume{810},
\bfpage{78}.
\doiurl{10.1088/0004-637X/810/1/78}.
\end{barticle}
\endbibitem

\bibitem[\protect\citeauthoryear{{McClintock}, {Norton}, and
  {Li}}{2014}]{McClintock2014}
\begin{barticle}
\bauthor{\bsnm{{McClintock}}, \binits{B.H.}},
\bauthor{\bsnm{{Norton}}, \binits{A.A.}},
\bauthor{\bsnm{{Li}}, \binits{J.}}:
\byear{2014},
\batitle{{Re-examining Sunspot Tilt Angle to Include Anti-Hale Statistics}}.
\bjtitle{\apj}
\bvolume{797},
\bfpage{130}.
\doiurl{10.1088/0004-637X/797/2/130}.
\end{barticle}
\endbibitem

\bibitem[\protect\citeauthoryear{{Miesch} and {Dikpati}}{2014}]{Miesch2014}
\begin{barticle}
\bauthor{\bsnm{{Miesch}}, \binits{M.S.}},
\bauthor{\bsnm{{Dikpati}}, \binits{M.}}:
\byear{2014},
\batitle{{A Three-dimensional Babcock-Leighton Solar Dynamo Model}}.
\bjtitle{\apjl}
\bvolume{785},
\bfpage{L8}.
\doiurl{10.1088/2041-8205/785/1/L8}.
\adsurl{http://esoads.eso.org/abs/2014ApJ...785L...8M}.
\end{barticle}
\endbibitem

\bibitem[\protect\citeauthoryear{{Mu{\~n}oz-Jaramillo}
  \textit{et~al.}}{2013}]{Munoz2013}
\begin{barticle}
\bauthor{\bsnm{{Mu{\~n}oz-Jaramillo}}, \binits{A.}},
\bauthor{\bsnm{{Dasi-Espuig}}, \binits{M.}},
\bauthor{\bsnm{{Balmaceda}}, \binits{L.A.}},
\bauthor{\bsnm{{DeLuca}}, \binits{E.E.}}:
\byear{2013},
\batitle{{Solar Cycle Propagation, Memory, and Prediction: Insights from a
  Century of Magnetic Proxies}}.
\bjtitle{\apjl}
\bvolume{767},
\bfpage{L25}.
\doiurl{10.1088/2041-8205/767/2/L25}.
\end{barticle}
\endbibitem

\bibitem[\protect\citeauthoryear{{Mu{\~n}oz-Jaramillo}
  \textit{et~al.}}{2015}]{Munoz2015}
\begin{barticle}
\bauthor{\bsnm{{Mu{\~n}oz-Jaramillo}}, \binits{A.}},
\bauthor{\bsnm{{Senkpeil}}, \binits{R.R.}},
\bauthor{\bsnm{{Windmueller}}, \binits{J.C.}},
\bauthor{\bsnm{{Amouzou}}, \binits{E.C.}},
\bauthor{\bsnm{{Longcope}}, \binits{D.W.}},
\bauthor{\bsnm{{Tlatov}}, \binits{A.G.}},
\bauthor{\bsnm{{Nagovitsyn}}, \binits{Y.A.}},
\bauthor{\bsnm{{Pevtsov}}, \binits{A.A.}},
\bauthor{\bsnm{{Chapman}}, \binits{G.A.}},
\bauthor{\bsnm{{Cookson}}, \binits{A.M.}},
\bauthor{\bsnm{{Yeates}}, \binits{A.R.}},
\bauthor{\bsnm{{Watson}}, \binits{F.T.}},
\bauthor{\bsnm{{Balmaceda}}, \binits{L.A.}},
\bauthor{\bsnm{{DeLuca}}, \binits{E.E.}},
\bauthor{\bsnm{{Martens}}, \binits{P.C.H.}}:
\byear{2015},
\batitle{{Small-scale and Global Dynamos and the Area and Flux Distributions of
  Active Regions, Sunspot Groups, and Sunspots: A Multi-database Study}}.
\bjtitle{\apj}
\bvolume{800},
\bfpage{48}.
\doiurl{10.1088/0004-637X/800/1/48}.
\end{barticle}
\endbibitem

\bibitem[\protect\citeauthoryear{{Petrie}, {Petrovay}, and
  {Schatten}}{2014}]{Petrie2014}
\begin{barticle}
\bauthor{\bsnm{{Petrie}}, \binits{G.J.D.}},
\bauthor{\bsnm{{Petrovay}}, \binits{K.}},
\bauthor{\bsnm{{Schatten}}, \binits{K.}}:
\byear{2014},
\batitle{{Solar Polar Fields and the 22-Year Activity Cycle: Observations and
  Models}}.
\bjtitle{\ssr}
\bvolume{186},
\bfpage{325}.
\doiurl{10.1007/s11214-014-0064-4}.
\end{barticle}
\endbibitem

\bibitem[\protect\citeauthoryear{{Petrovay}}{2010}]{Petrovay2010}
\begin{barticle}
\bauthor{\bsnm{{Petrovay}}, \binits{K.}}:
\byear{2010},
\batitle{{Solar Cycle Prediction}}.
\bjtitle{\lrsp}
\bvolume{7},
\bfpage{6}.
\doiurl{10.12942/lrsp-2010-6}.
\end{barticle}
\endbibitem

\bibitem[\protect\citeauthoryear{{Schatten}
  \textit{et~al.}}{1978}]{Schatten1978}
\begin{barticle}
\bauthor{\bsnm{{Schatten}}, \binits{K.H.}},
\bauthor{\bsnm{{Scherrer}}, \binits{P.H.}},
\bauthor{\bsnm{{Svalgaard}}, \binits{L.}},
\bauthor{\bsnm{{Wilcox}}, \binits{J.M.}}:
\byear{1978},
\batitle{{Using dynamo theory to predict the sunspot number during solar cycle
  21}}.
\bjtitle{\grl}
\bvolume{5},
\bfpage{411}.
\doiurl{10.1029/GL005i005p00411}.
\end{barticle}
\endbibitem

\bibitem[\protect\citeauthoryear{{Svalgaard}, {Cliver}, and
  {Kamide}}{2005}]{Svalgaard2005}
\begin{barticle}
\bauthor{\bsnm{{Svalgaard}}, \binits{L.}},
\bauthor{\bsnm{{Cliver}}, \binits{E.W.}},
\bauthor{\bsnm{{Kamide}}, \binits{Y.}}:
\byear{2005},
\batitle{{Sunspot cycle 24: Smallest cycle in 100 years?}}
\bjtitle{\grl}
\bvolume{32},
\bfpage{L01104}.
\doiurl{10.1029/2004GL021664}.
\end{barticle}
\endbibitem

\bibitem[\protect\citeauthoryear{{Tlatov} and {Pevtsov}}{2014}]{Tlatov2014}
\begin{barticle}
\bauthor{\bsnm{{Tlatov}}, \binits{A.G.}},
\bauthor{\bsnm{{Pevtsov}}, \binits{A.A.}}:
\byear{2014},
\batitle{{Bimodal Distribution of Magnetic Fields and Areas of Sunspots}}.
\bjtitle{\solphys}
\bvolume{289},
\bfpage{1143}.
\doiurl{10.1007/s11207-013-0382-9}.
\end{barticle}
\endbibitem

\bibitem[\protect\citeauthoryear{{Toriumi} \textit{et~al.}}{2017}]{Toriumi2017}
\begin{barticle}
\bauthor{\bsnm{{Toriumi}}, \binits{S.}},
\bauthor{\bsnm{{Schrijver}}, \binits{C.J.}},
\bauthor{\bsnm{{Harra}}, \binits{L.K.}},
\bauthor{\bsnm{{Hudson}}, \binits{H.}},
\bauthor{\bsnm{{Nagashima}}, \binits{K.}}:
\byear{2017},
\batitle{{Magnetic Properties of Solar Active Regions That Govern Large Solar
  Flares and Eruptions}}.
\bjtitle{\apj}
\bvolume{834},
\bfpage{56}.
\doiurl{10.3847/1538-4357/834/1/56}.
\end{barticle}
\endbibitem

\bibitem[\protect\citeauthoryear{{Upton} and {Hathaway}}{2014}]{Upton2014}
\begin{barticle}
\bauthor{\bsnm{{Upton}}, \binits{L.}},
\bauthor{\bsnm{{Hathaway}}, \binits{D.H.}}:
\byear{2014},
\batitle{{Effects of Meridional Flow Variations on Solar Cycles 23 and 24}}.
\bjtitle{\apj}
\bvolume{792},
\bfpage{142}.
\doiurl{10.1088/0004-637X/792/2/142}.
\adsurl{http://esoads.eso.org/abs/2014ApJ...792..142U}.
\end{barticle}
\endbibitem

\bibitem[\protect\citeauthoryear{{van Ballegooijen} and
  {Choudhuri}}{1988}]{vanBallegooijen1988}
\begin{barticle}
\bauthor{\bsnm{{van Ballegooijen}}, \binits{A.A.}},
\bauthor{\bsnm{{Choudhuri}}, \binits{A.R.}}:
\byear{1988},
\batitle{{The possible role of meridional flows in suppressing magnetic
  buoyancy}}.
\bjtitle{\apj}
\bvolume{333},
\bfpage{965}.
\doiurl{10.1086/166805}.
\end{barticle}
\endbibitem

\bibitem[\protect\citeauthoryear{{Wang} and {Sheeley}}{1991}]{WangSheeley1991}
\begin{barticle}
\bauthor{\bsnm{{Wang}}, \binits{Y.-M.}},
\bauthor{\bsnm{{Sheeley}}, \binits{N.R.} \bsuffix{Jr.}}:
\byear{1991},
\batitle{{Magnetic flux transport and the sun's dipole moment - New twists to
  the Babcock-Leighton model}}.
\bjtitle{\apj}
\bvolume{375},
\bfpage{761}.
\doiurl{10.1086/170240}.
\end{barticle}
\endbibitem

\bibitem[\protect\citeauthoryear{{Wang}, {Nash}, and
  {Sheeley}}{1989}]{WangSheeley1989}
\begin{barticle}
\bauthor{\bsnm{{Wang}}, \binits{Y.-M.}},
\bauthor{\bsnm{{Nash}}, \binits{A.G.}},
\bauthor{\bsnm{{Sheeley}}, \binits{N.R.} \bsuffix{Jr.}}:
\byear{1989},
\batitle{{Magnetic flux transport on the sun}}.
\bjtitle{Science}
\bvolume{245},
\bfpage{712}.
\doiurl{10.1126/science.245.4919.712}.
\end{barticle}
\endbibitem

\bibitem[\protect\citeauthoryear{{Yeates} and
  {Mu{\~n}oz-Jaramillo}}{2013}]{Yeates2013}
\begin{barticle}
\bauthor{\bsnm{{Yeates}}, \binits{A.R.}},
\bauthor{\bsnm{{Mu{\~n}oz-Jaramillo}}, \binits{A.}}:
\byear{2013},
\batitle{{Kinematic active region formation in a three-dimensional solar dynamo
  model}}.
\bjtitle{\mnras}
\bvolume{436},
\bfpage{3366}.
\doiurl{10.1093/mnras/stt1818}.
\adsurl{http://esoads.eso.org/abs/2013MNRAS.436.3366Y}.
\end{barticle}
\endbibitem

\bibitem[\protect\citeauthoryear{{Yeates}, {Baker}, and {van
  Driel-Gesztelyi}}{2015}]{Yeates2015}
\begin{barticle}
\bauthor{\bsnm{{Yeates}}, \binits{A.R.}},
\bauthor{\bsnm{{Baker}}, \binits{D.}},
\bauthor{\bsnm{{van Driel-Gesztelyi}}, \binits{L.}}:
\byear{2015},
\batitle{{Source of a Prominent Poleward Surge During Solar Cycle 24}}.
\bjtitle{\solphys}
\bvolume{290},
\bfpage{3189}.
\doiurl{10.1007/s11207-015-0660-9}.
\end{barticle}
\endbibitem

\end{thebibliography}

\end{article}
\end{document}